\documentclass[aps,pre,reprint,superscriptaddress,floatfix,twocolumn,10pt]{revtex4-2}
\usepackage[english]{babel}
\usepackage{amsmath,bm}
\usepackage{amssymb}
\usepackage{enumitem}
\usepackage{comment}
\usepackage{setspace}
\usepackage{tabularx}
\usepackage{microtype}
\usepackage{rotating}
\usepackage{amstext}
\usepackage{dcolumn}

\usepackage{graphicx}
\usepackage[bottom]{footmisc}
\usepackage{multirow}
\usepackage{array}
\usepackage{caption}
\usepackage{rotating}
\usepackage{keyval}
\usepackage{ifthen}
\usepackage{chemfig}
\usepackage{lmodern} 
\usepackage{mathtools}
\usepackage{tabularx}
\usepackage{braket}
\usepackage{subfig} 
\usepackage{siunitx}
\usepackage[font=small]{caption}
\usepackage{subeqnarray}
\includecomment{reprint}
\excludecomment{preprint}
\usepackage{xcolor}

\definecolor{dgreen}{rgb}{0,.5,0}
\definecolor{dred}{rgb}{.7,.0,.0}
 
\def\XXint#1#2#3{{\setbox0=\hbox{$#1{#2#3}{\int}$}
     \vcenter{\hbox{$#2#3$}}\kern-.5\wd0}}

\newcommand{\indiceGauche}[2]{{\vphantom{#2}}_{#1}#2}

\newcommand{\br}{\mathbf{r}}
\newcommand{\bu}{\mathbf{u}}

\newcommand{\bA}{\mathbf{A}}

\newcommand{\fT}{\mathfrak{T}}

\newcommand{\fL}{\mathfrak{L}}

\newcommand{\fn}{\mathfrak{n}}
\newcommand{\fS}{\mathfrak{S}}

\newcommand{\bn}{\mathbf{n}}

\newcommand{\bh}{\mathbf{h}}
\newcommand{\bv}{\mathbf{v}}
\newcommand{\bM}{\mathbf{M}}
\newcommand{\bT}{\mathbf{T}}

\newcommand{\beps}{\boldsymbol{\epsilon}}

\usepackage{floatrow}
\usepackage[colorlinks,bookmarks=false,citecolor=blue,linkcolor=red,urlcolor=blue]{hyperref}

\begin{document}

\title{Theory of Rayleigh molecular light scattering by isotropic polar fluids revisited}

\author{Pierre-Michel D\'{e}jardin}
\affiliation{Laboratoire de Mod\'{e}lisation Pluridisciplinaire et Simulations, Universit\'{e} de Perpignan Via Domitia, 52 avenue Paul Alduy, F-66860 Perpignan, France}

\begin{abstract}
The molecular theory of Rayleigh light scattering in dense isotropic polar fluids is reconsidered by suitably adapting local field concepts of electrostatics to propagating electromagnetic waves, hence accounting for both the rotational and dipole-induced dipole (DID) contributions. Simple analytical equations are derived for the various Rayleigh ratios relevant to lateral light scattering in various situations, namely pure DID, pure rotations and mixed contributions. For pure DID, the derived Rayleigh ratios are entirely analytical and very simple, while for pure rotation, the use of rotational mean field approximation is justified, hence allowing the description of Rayleigh ratios in terms of a single orientational correlation parameter $g_2^{VH}$ that is straightforwardly determined as a positive root of a quadratic algebraic equation. Simple expressions for the Rayleigh ratios are also derived in two mixed situations where DID dominates rotation and when rotation dominates DID. The analytical formulas are directly compared with experimental data concerning five liquids, namely carbon tetrachloride, benzene, toluene, carbon disulfide and nitrobenzene. The theoretical formulas show excellent agreement, not only with all the Rayleigh ratios across the whole visible wavelength range, but also for the derivative of the squared refractive index with respect to number density times number density. This allows to demonstrate that the scaling of the Rayleigh ratios with the internal field factor $\fL$ is generally not trivial, and that $\fL$ precise knowledge is \textit{not required} as far as light scattering is concerned. The theory so dealt with also explains why in dilute situations, the molecular polarizability anisotropy deduced from light scattering data agrees with quantum ab-initio calculations, while being slightly smaller due to orientational correlations whenever the latter are relevant. Relation to previous theoretical work is also discussed. 
\end{abstract}

\maketitle

\section{Introduction}\label{1}
Despite enormous theoretical and experimental efforts in trying to interpret experimental data regarding light scattering by liquids in molecular terms \cite{Dezelic1970PureApplChem}, the detailed analysis of the various experimental light scattered intensities still nowadays remains complicated. This may, on one hand, be attributed to the fact that the various intensities are never given in absolute units, but also on the other hand to the lack of existence of correct analytical formulas for the so-called Rayleigh ratios (differential cross-sections per unit volume) expressed in molecular terms. The important discovery by McTague and co-workers \cite{McTague1968PRL,McTague1969PRL} is that the light scattering signal is \textit{never} fully polarized, had this signal arisen either from a liquid, either from a gas, and had the constituents of the liquid or gas (atoms or molecules) been anisotropically polarizable or not. Accounting for this phenomenon, nowadays explained by the so-called dipole-induced dipole (DID) mechanism \cite{BernePecora1970Book}, largely complicates the derivation of such formulas ; nor the Smoluchowski-Einstein theory \cite{Smoluchowski1908AnnPhys,Einstein1910AnnPhys}, nor the more sophisticated Ornstein-Zernike one \cite{OZ1926ZPhys} are able to explain the systematic presence of depolarized scattered light by polar fluids in molecular terms.

The reason for partial depolarization of scattered light may be traced back to the work of Gans \cite{Gans1921AnnPhys}. By considering anisotropically polarizable small scatterers in the ideal gas phase, this author was able to demonstrate that it is because of this polarizability anisotropy property (and therefore because of the rotational motion of the scatterers \cite{BernePecora1970Book}) that a depolarized component of scattered light may arise. This key result suggests that in order to have a theoretical explanation of the experimental results of McTague et al. \cite{McTague1968PRL,McTague1969PRL}, some "molecular" anisotropy must be induced by the incident light beam (and quite remarkably, this was conjectured by Kielich \cite{Kielich1967JCP} one year before experimental evidence). Hence it becomes obvious that an effect of importance has been ignored in the derivations of Einstein, and Ornstein and Zernike. The "source" of anisotropy for isotropically polarizable scatterers cannot be found in the intrinsic properties of the scatterers themselves. Rather, such incident wave-induced anisotropy is contained in the specific interaction term in the expression for the local field (see Born and Wolf \cite{BornWolf2005Book} , Chapter 2, first equations of Section 2.4), giving rise to the so-called dipole-induced dipole (DID) contribution to light scattering alluded to earlier in the text. Moreover, contrarily to the electrostatic situation where this specific part may be ignored in the dilute situation \cite{Bottcher1973Book}, the experimental results of McTague et al. strongly suggest that this is \textit{never} possible at optical frequencies. In the dilute situation, molecular polarizability anisotropies may be extracted from experimental data using the Gans formula for the depolarization ratio \cite{Gans1921AnnPhys,Kerker2013Book}, where in this formula and in Gans's theory in general, the specific interaction term in the local field \textit{is not accounted for at all}. Moreover, as remarked quite recently by Pabst and Blochowicz \cite{Pabst2022JCP} these experimental values of the molecular polarizability anisotropy agree \textit{in all respects} with quantum ab-initio calculations in the large majority of cases, and it is still not obvious \textit{how this is possible} \cite{Keyes2024JCP,Pabst2024JCP}. In principle, a definite answer to this question should involve a full quantum-mechanical treatment. However, Gelbart \cite{Gelbart1974Book} has shown that this is, most of the time, not necessary and that the local field approach (involving only a classical treatment) is, to a large extent, equivalent to a full quantum one. It follows that in the context of the local field approach, a molecular theory of light scattering must account for both molecular rotation and DID contributions. Here, it is worthwhile quickly reviewing the evolution of the subject after Einstein's and Ornstein and Zernike's contributions.

In effect, the investigations in the subject of light scattering evolved slowly, because of various uncertainties regarding the measurements. Clearly, the measured light scattered intensities were far too low in order to expect new theoretical predictions. In this respect, Brice et al. \cite{Brice1950JOptSocAm} proposed a new setup (the so-called Brice-Phoenix setup, see the paper by these authors and Reference \citenum{Dezelic1970PureApplChem} for a review) which allowed for the first time to measure \textit{absolute} scattered intensities. The Einstein theory lacked a strong molecular basis at the time, although it was felt by researchers that the derivation of his formula could be accomplished via the statistical-mechanical route. The first author who attempted to accomplish this task was Fixman \cite{Fixman1955JCP} (Einstein's results was also rederived twenty years later by Felderhof \cite{Felderhof1974Physica} using an original method), while some years later, Coumou \cite{Coumou1960JcollSci} and Coumou et al. \cite{Coumou1964TransFarSoc}, following an intuition of Cabannes \cite{Kerker2013Book}, experimentally showed that the Rayleigh ratio $R_{iso}$ corresponding to isotropically scattered light can be related to the refractive index of the fluid $\fn$ by Einstein's equation \cite{Einstein1910AnnPhys}, viz.
\begin{eqnarray}
    R_{iso}=\frac{\pi^2}{2\rho_0\lambda_0^4}\fS\left(\rho_0\frac{\partial\fn^2}{\partial\rho_0}\right)_T^2\label{RisoEinstein}
\end{eqnarray}
where $\rho_0$ is the number density, $\lambda_0$ is the wavelength of the incident light beam, $\fS$ is the zero wavevector limit of the structure factor of the fluid \cite{Hansen2006Book} given by
\begin{eqnarray}
    \fS=\rho_0kT\chi_T
    \label{Sstatic}
\end{eqnarray}
$k$ being Boltzmann's constant, $T$ the absolute temperature and $\chi_T$ the isothermal compressibility of the fluid. Indeed, Eq.\eqref{RisoEinstein} is valid if the temperature variation of $\fn$ is neglected. 

Now, importantly, Coumou et al. \cite{Coumou1964TransFarSoc} \textit{experimentally} demonstrated that Eq.\eqref{RisoEinstein} holds to a high degree of accuracy at liquid densities, normal pressures and room temperature, had the scatterers been \textit{isotropically or anisotropically polarizable} (CCl$_4$ and C$_6$H$_6$ are notorious examples of both kinds of -molecular- scatterers). We also note that at the time, it was moreover known  that despite being isotropically polarizable, light scattered by liquid CCl$_4$ shows a substantial depolarized scattered component \cite{Coumou1964TransFarSoc2,Dezelic1970PureApplChem}, in contradiction with the existing molecular theory which was predicting fully polarized scattered light for isotropically polarizable scatterers. Besides, Kielich \cite{Kielich1965ActaPhysPol} and Pecora and Steele \cite{PecoraSteele1965JCP} established a number of relations between scattered intensities (therefore, between Rayleigh ratios) corresponding to various polarizations of the incident and scattered waves, and this for arbitrary scattering angle (see again Reference \citenum{Dezelic1970PureApplChem} where these relations are stated in a very simple fashion). These relations are important in that they do not rely on any molecular model, so that they can routinely be used in order to analyze any light scattering experiment, in particular when the scattered light arises from an isotropic polar fluid. They moreover show that the Rayleigh ratio of unpolarized incident and scattered light is the sum of an isotropic component $R_{iso}$ and an anisotropic component $R_{an}$. By attributing depolarized scattered light to fluctuations of the off-diagonal elements of the linear dielectric permittivity tensor (thereby following Einstein), Dezelic and Vavra \cite{DezelicVavra1966CroatChemActa} were able to relate $R_{an}$ to $(\rho_0\partial_{\rho_0}\fn^2)_T$ for anisotropically polarizable scatterers. Their equation is, for $90^{\circ}$ light scattering (also termed lateral light scattering)
\begin{eqnarray}
    R_{an}=\frac{13\pi^2\kappa^2G}{10\rho_0\lambda_0^4} \left(\rho_0\frac{\partial\fn^2}{\partial\rho_0}\right)_T^2
    \label{DezelicVavraRan}
\end{eqnarray}
where $G$ is an orientational correlation factor related to pair ordering of long molecular axes \cite{Benoit1957JPhysRadium,Bottcher1978Book} and $\kappa$ is the relative molecular polarizability anisotropy. For symmetric top molecules, $\kappa$ is defined by
\begin{eqnarray}
    \kappa=\frac{\Delta\alpha}{3\bar{\alpha}}
    \label{kappaDef}
\end{eqnarray}
$\Delta\alpha$ is the natural molecular polarizability anisotropy, viz.
\begin{eqnarray}
\nonumber
    \Delta\alpha=\alpha_{\vert\vert}-\alpha_{\perp},
\end{eqnarray}
while $\bar\alpha$ is the average molecular anisotropy, i.e.,
\begin{eqnarray}
\nonumber
    \bar{\alpha}=\frac{\alpha_{\vert\vert}+2\alpha_{\perp}}{3},
\end{eqnarray}
and in the last two equations $\alpha_{\vert\vert}$ is the natural (i.e. in the ideal gas phase) scalar molecular polarizability along the long molecular axis, $\alpha_{\perp}$ that perpendicular to the long axis. We remark in passing that Eq.\eqref{DezelicVavraRan} can only be applied if $\kappa$ noticeably differs from zero. Leaving this fact apart for the moment, the model-independent relations between Rayleigh ratios valid for any scattering angle are completely specified by the knowledge of $R_{iso}$ and $R_{an}$, so that at first glance, a molecular (i.e. microscopic) theory of light scattering can be obtained if $(\rho_0\partial_{\rho_0}\fn^2)_T$ is specified in molecular terms. Equally stated, one needs to express $\fn$ in terms of density $\rho_0$ and individual molecular (atomic) characteristics, in other words, to find an optical dielectric equation of state that is valid from gaseous to liquid densities. Using the Lorenz-Lorentz equation, Einstein derived \cite{Einstein1910AnnPhys}
\begin{eqnarray}
\nonumber
        \left(\rho_0\frac{\partial\fn^2}{\partial\rho_0}\right)_T^2&=&(\fn^2-1)^2\fL_L^2(\lambda_0)\\
        &=&16\pi^2\rho_0^2\bar{\alpha}^2\fL_L^4(\lambda_0).
        \label{Einstein2}
    \end{eqnarray}
where $\fL_L$ is the so-called Lorentz internal field factor \cite{Bottcher1973Book,Bottcher1978Book}, viz.
\begin{eqnarray}
    \nonumber
    \fL_L(\lambda_0)=\frac{\fn^2(\lambda_0)+2}{3}.
\end{eqnarray}
This result, together with Eqs.\eqref{RisoEinstein} and \eqref{DezelicVavraRan} have lead numerous researchers to state $\fL^4$ scaling of the light scattering intensities and Rayleigh ratios (see References \citenum{MalmbergLippincott1968JcollIntSci,Burnham75,Patterson1975JCP,KeyesLadanyi1984Book} and References therein, to name but a few), where now $\fL$ is the internal field factor that always occurs in local field theory (this may of course differ from $\fL_L$) \cite{Bottcher1973Book,Bottcher1978Book}. It was nevertheless demonstrated experimentally that the right hand side of Eq.\eqref{Einstein2} \textit{overestimates} $(\rho_0\partial_{\rho_0}\fn^2)_T$ by $5-10\%$ \cite{Coumou1964TransFarSoc,Kerker2013Book,Beysens1977JCP}, leading to an overestimate of light scattered intensities by $10-20\%$ \cite{Dezelic1970PureApplChem}, i.e. a \textit{large} factor. Moreover, it was also shown that the Kirkwood and Yvon theories of the refractive index \cite{Kirkwood1936JCP,Yvon1937Book} lead to numerical values of $(\rho_0\partial_{\rho_0}\fn^2)_T$ in \textit{worse} agreement with experimental data than the result provided by the Lorenz-Lorentz equation. Thus, optical dielectric equations of state of empirical origin were used, which have their own pros and cons \cite{Kerker2013Book}, but cannot be quoted as completely satisfactory from a theoretical point of view.

It is important to recall that the correlation factor $G$ is currently interpreted as follows. If $G<1$, pairs of long molecular axes tend to orient perpendicular, $G>1$ means a trend to parallel alignment while if $G=1$, no orientation is preferred (This value is also expected in the ideal gas phase or in dilute situations). The so-called vertical depolarization ratio $r_v$ \cite{Kerker2013Book,Dezelic1970PureApplChem}, given in molecular terms by
\begin{eqnarray}
    r_v=\frac{3\kappa^2G}{5\fS+4\kappa^2G}
    \label{rvG}
\end{eqnarray}
leads \textit{invariably}, if the gas phase $\kappa$ value is used, to $G<1$ for virtually \textit{all} molecular systems at room temperature (for Benzene, $0.53<G<0.58$) and \textit{seems} sensitive to the value of the probing wavelength, \textit{in contradiction with its theoretical definition}. If the molecules are polar and linear (and where the permanent dipole is along the long axis), this, at first glance, contradicts the Kirkwood-Fröhlich theory of the dielectric constant of isotropic polar fluids \cite{Kirkwood1939JCP,Frohlich1958Book,Bottcher1973Book}. 

Now, both in the dilute situation and in the vapor phase, $G=1$ and $\fS\approx 1$, and quite remarkably, in both situations, Eq.\eqref{rvG} together with $\fL^2$ scaling of the \textit{depolarized} light scattering intensity lead, as remarked by Pabst and Blochowicz \cite{Pabst2022JCP}, to $\kappa$ values in \textit{perfect agreement} with quantum ab-initio calculations of the molecular polarizability tensor for circa 100 molecular liquids, in contradiction with $\fL^4$ scaling in all situations. Moreover, Eq.\eqref{rvG}, at least when $G=1$ does not contain DID effects, in apparent contradiction with the experimental results of McTague et al. \cite{McTague1968PRL,McTague1969PRL} where substantial depolarization of scattered light occurs (and therefore, DID cannot be excluded since this occurs for CCl$_4$) \textit{even in the gaseous phase}. 

In order to remove the apparent inconsistencies of the $G$ values provided by Eq.\eqref{rvG} and to recover the $\fL^4$ scaling of all light scattering intensities, Burnham et al.\cite{Burnham75} suggested that even in the dilute situations, $\kappa$ should be replaced by $\kappa_e<\kappa$ an effective polarizability anisotropy (so that in the expression $\kappa_e^G$, $G$ can approach $1$). In order to justify this, they argue that the DID effect manifests itself in the light scattered intensities by an $\fL^4$ scaling of the \textit{rotational} component of \textit{dynamic light scattering} onto which a background attributed to a broad line is superimposed, the latter background being in turn attributed to collision-induced scattering. It follows that in Burnham et al.'s picture, the DID effect is shared in between a rotational component and a collision-induced term. The lineshape of the broad line is guessed from the wings of the dynamic light scattering spectrum, fitted by a Lorentzian and subtracted from the total spectrum. The result of this operation is then integrated over frequencies in order to effectively single out the rotational component of the depolarized light scattering intensity, so that the rotational contribution to the Rayleigh ratio of depolarized scattered light $R_{VH}^{r}$ is written as \cite{Burnham75,Battaglia1979MolPhys}
\begin{eqnarray}
    R_{VH}^{r}=\left[\frac{2\pi}{\lambda_0}\right]^4\fL^4(\lambda_0)\frac{\rho_0[\Delta\alpha_e(\lambda_0)]^2}{15}G
    \label{RVHSubtract}
\end{eqnarray}
In practice, $\fL$ is replaced by the Lorentz internal field factor in this equation. This procedure apparently restores "reasonable" values of $G$, as argued by Burnham et al.\cite{Burnham75} and some time later by Battaglia et al.\cite{Battaglia1979MolPhys}. By extending Felderhof's and Fixman's theories of light scattering to anisotropically polarizable molecules \cite{KeyesLadanyi1984Book}, Keyes and Ladanyi provided an analytical expression for $\Delta\alpha_e(\lambda_0)$ in terms of the natural $\Delta\alpha$ \cite{KeyesLadanyi1977MolPhys3}. In terms of $\kappa$, $\kappa_e$ $\bar\alpha$ and an effective mean polarizability $\Delta\alpha_e=3\kappa_e\bar\alpha_e$, this polarizability anisotropy is essentially given, using our notations, by
\begin{eqnarray}
   \bar\alpha_e(\lambda_0)\kappa_e(\lambda_0)=\bar\alpha\kappa-\bar\alpha\left[\frac{1+\kappa}{2\kappa}-\kappa\right]\left[1-\frac{\fL_L(\lambda_0)}{\fL(\lambda_0)}\right]
    \label{KappaeBL}
\end{eqnarray}
where the true internal field factor $\fL$ is specified by Eq.\eqref{OpticalDielectricEqOfState} (see later in the text). Furthermore, they found through their quite complicated calculations a justification of Eq.\eqref{RVHSubtract}. Nevertheless, if the Lorentz field factor is used in Eq.\eqref{RVHSubtract} and \eqref{KappaeBL} then $\Delta\alpha_e=\Delta\alpha$. Furthermore, when $\kappa=0$, then \textit{necessarily}, $\fL=\fL_L$ and $\kappa_e=0$, otherwise, Eq.\eqref{KappaeBL} is singular when $\kappa=0$. In other words, since the collision-induced term which must be superimposed on Eq.\eqref{RVHSubtract} is unknown, the Keyes-Ladanyi theory cannot, in practice, be applied to fluids made of isotropically polarizable molecules. This is because the removal of the singularity occurring in Eq.\eqref{KappaeBL} must a priori be sought in the collision-induced term, and the Keyes-Ladanyi theory is still currently at a much too formal stage in order to allow an explicit calculation of the latter. Here, numerical simulations of the Keyes-Ladanyi theory cannot help since a very small $\kappa$ value inserted in Eq.\eqref{KappaeBL} may lead to severe numerical drifts. At last, back to $\kappa\neq0$, experimental data show that the so-called background which is subtracted from the total spectrum may represent quite a large contribution to the total depolarized light scattering ratio $R_{VH}$ (more than $40\%$ for C$_6$H$_6$ and CS$_2$ \cite{Battaglia1979MolPhys}). Since the background is subtracted by a fitting procedure, it may legitimately be suspected that a non-negligible fraction of the scattered signal arising from the sample is subtracted in an undesirable way. Pabst and Blochowicz's recent work \cite{Pabst2022JCP} implicitly suggest this, re-rising this quite old and never-ending debate of Rayleigh ratio scaling again \cite{Keyes2024JCP,Pabst2024JCP}. Here, one criticism which can be made to Eq.\eqref{RVHSubtract} is as follows. Independent of the expression of the internal field factor $\fL$, it can nevertheless be shown that the optical dielectric equation of state for isotropic polar fluids can always be written as follows
\begin{eqnarray}
    \fn^2-1=4\pi\rho_0\fL\bar\alpha
    \label{OpticalDielectricEqOfState}
\end{eqnarray}
which entails that $R_{VH}^{r}$ values as rendered by Eq.\eqref{RVHSubtract} \textit{might be dependent of the shape of the cavity}, because it implicitly means that the light scattered intensity is divided through by the square of the Maxwell field (and not that of the vacuum field). In other words, changing $\fL$ \textit{changes} the values rendered by $R_{VH}^{r}$, and despite $\fL_L$ is quite well justified theoretically, other expressions for $\fL$ exist (particularly empirical ones) \cite{Bottcher1978Book,Kerker2013Book} that agree \textit{much better} with experimental $(\rho_0\partial_{\rho_0}\fn^2)_T$ than that provided by the Lorenz-Lorentz equation. In the Keyes-Ladanyi theory, Rayleigh ratios are also obtained by dividing the light scattered intensity by the squared Maxwell field amplitude. In effect, in this context, the specimen occupies \textit{the whole space} including the detector \cite{Frenkel1980JCP} and necessarily, the light source and the whole measuring device. Burnham et al. justified the concept of an "experimental $\kappa$" by attaching themselves to show that the Onsager-Scholte local field model \cite{Onsager1936JACS,Scholte1949Physica} involving \textit{ellipsoidal} cavities containing \textit{one} molecule only are most adequate for interpreting light scattering data. Of course, in Onsager-like theories, one always have $G=1$ \cite{Bottcher1973Book}. It nevertheless forces one to multiply Rayleigh ratios calculated by the statistical mechanical route by $\fL^2$ again \cite{Burnham75,Patterson1975JCP} in order to agree with Einstein's thermodynamic fluctuation theory since it is exactly in this proportion that the two calculations \textit{seemingly} agree. This is however difficult to accept as a theoretical argument, as a spherical cavity should work equally well for calculating the Rayleigh ratios. It follows that in our opinion, $R_{VH}^{r}$ as defined by Eq.\eqref{RVHSubtract} is \textit{not an intensive quantity}. Therefore \textit{stricto sensu}, this equation \textit{cannot} be used in order to represent a light scattering characteristics of the system, despite apparent agreement with experiment (we discuss this also in Section \ref{7}). In addition, as argued recently \cite{Pabst2024JCP}, in subtracting a background (and this operation always has some degree of arbitrariness), one may in effect subtract a non-negligible part of the signal arising from the sample itself. It follows that several ingredients are lacking in the molecular theory of light scattering, the first of these being analytical formulas for the Rayleigh ratios in the liquid state for isotropically polarizable point scatterers. In effect, such formulas would help in concluding further regarding scaling of the Rayleigh ratios with $\fL$. Unfortunately, there is presently no simple derived formula describing the mixed situations where both DID and rotation contribute to light scattering. Finally, the sole current way of removing the low (and wavelength-dependent) $G$ values predicted by Eq.\eqref{rvG} being subtracting a background of arbitrary nature makes us feel that orientational correlation factor/factors pertaining to molecular light scattering by isotropic polar fluids needs/need to be theoretically redefined \textit{for this experiment}. 

Owing to what precedes, it is the purpose of this work to entirely revisit the theory of Rayleigh molecular light scattering by isotropic polar fluids in the context of local field theory. The optical dielectric equation of state \eqref{OpticalDielectricEqOfState} together with an integral formula valid in the first Born approximation \cite{Born1926ZPhys,Morse1953Book} for the Rayleigh ratio with arbitrary incident and scattered wave polarizations can straightforwardly be derived using elements of radiation theory combined with local field concepts that are easily transposed to propagation of electromagnetic waves in continuous media. Here, for lateral light scattering, we first derive generic formulas in Section \ref{2} for $R_{VV}$, $R_{HH}$ and $R_{VH}=R_{HV}$, which are the Rayleigh ratios for vertically polarized incident and scattered waves ($VV$ index),  horizontally polarized incident and scattered waves ($HH$ index), and vertically polarized incident and horizontally polarized scattered waves and vice-versa ($VH$ and $HV$ indexes) respectively. For these Rayleigh ratios, the notations of Berne and Pecora \cite{BernePecora1970Book} are preferred over the Krishnan ones (see Reference \citenum{Dezelic1970PureApplChem} for a description of these notations) as we believe them more intuitive. Moreover, we define orientational correlation factors in Fröhlich's manner \cite{Frohlich1958Book} (i.e. orientational average of a quantity in the liquid phase divided by the orientational average of the same quantity in the ideal gas phase) so that, \textit{a priori}, three of them arise, one for each Rayleigh ratio,  which reduce to a single one, $g_{2}^{VH}$ in some circumstances. Then, three special situations are considered in Sections \ref{3}-\ref{5}, namely full DID, full rotation, and mixed contributions to light scattering, the first two situations being used as benchmarks for the more complicated third one. Since the theory does not contain \textit{any adjustable parameter}, it is directly compared with experimental data concerning liquid CCl$_4$ (carbon tetrachloride), C$_6$H$_6$ (Benzene), C$_7$H$_8$ (Toluene), CS$_2$ (carbon disulfide) and C$_6$H$_5$NO$_2$ (Nitrobenzene) in Section \ref{6}. At last, our theoretical findings are compared with previous results in Section \ref{7}. Since only $90^{\circ}$ light scattering is required to specify scattered intensities and Rayleigh ratios for any scattering angle \cite{Kielich1965ActaPhysPol,PecoraSteele1965JCP,Dezelic1970PureApplChem}, we restrict to this situation throughout the whole manuscript.

\begin{widetext}
\section{General formalism}\label{2}
We consider a plane progressive monochromatic electromagnetic wave propagating in a direction specified by the unit vector $\bn_0$. Reaching the specimen, and according to general scattering theory, the wave is isotropically scattered in all directions \cite{Jackson1999Book,BornWolf2005Book}. We detect scattered light in one propagating direction only, specified by a unit vector $\bn$, at a locus \textit{away} from the specimen. If $\bn_0$ and $\bn$ are not collinear, then these two unit vectors define a plane termed the scattering plane \cite{BernePecora1970Book}, and the scattering angle $\phi$ is defined by $\cos\phi=\bn\cdot\bn_0$. As already mentioned, it is enough to restrict to lateral light scattering so that $\phi=\pi/2$. Denoting by $\beps_0$ and $\beps$ unit vectors in the polarization directions of the incident and scattered waves, it is straightforward (but tedious, hence we do not provide a derivation here) starting from Maxwell's equations in matter to show that the Rayleigh ratio for these polarization states is, \textit{in the first Born approximation}, given by
\begin{eqnarray}
    \nonumber
    R_{\beps\beps_0}=\frac{\pi^2(\fn^2-1)^2}{\rho_0\lambda_0^4\bar\alpha^2}\Bigg[\indiceGauche{1}{\langle(\beps\cdot\bA(\bu_1)\cdot\beps_0)^2\rangle}_0+(\fS-1)\;\indiceGauche{2}{\langle(\beps\cdot\bA(\bu_1)\cdot\beps_0)(\beps\cdot\bA(\bu_2)\cdot\beps_0)\rangle}_0\Bigg]\\
    +\frac{(\fn^2-1)^4}{8\rho_0^2\lambda_0^4}\left(\int (\beps\cdot\bT(\br)\cdot\beps_0)^2g(\br)d\br\right)f_{\beps\beps_0}(\kappa),\qquad\qquad\qquad\qquad
    \label{ElasticRayleighFinal}
\end{eqnarray}
where $\bA(\bu)$ is the polarizability tensor of a molecule having the orientation of its long axis specified by unit vector $\bu$, viz.
\begin{eqnarray}
    \bA(\bu)=\bar{\alpha}(\mathbf{I}+\kappa\bM(\bu)),\qquad\bM(\bu)=3\bu\bu-\mathbf{I}
    \label{A}
\end{eqnarray}
$\bT(\br)$ is the dipole-dipole tensor here given by
\begin{eqnarray}
    \bT(\br)=\frac{3\hat{\br}\hat{\br}-\mathbf{I}}{r^3},\;\hat{\br}=\frac{\br}{r},\;\bT(\mathbf{0})=\mathbf{0}
    \label{T}
\end{eqnarray}
$g(\br)$ is the pair distribution function for separation of molecular centers specified by vector $\br$, that generally depends \textit{both} on $\br$ magnitude and orientations, the ellipticity factors $f_{\beps\beps_0}(\kappa)$ are formally given by (see Appendix \ref{AppA} for a description of these factors in terms of orientational averages)
\begin{eqnarray}
    f_{\beps\beps_0}(\kappa)=\frac{\indiceGauche{2}{\Bigg\langle\int (\beps\cdot\fT(\br,\bu_1,\bu_2)\cdot\beps_0)^2g(\br)d\br\Bigg\rangle}_0}{\bar{\alpha}^4\int(\beps\cdot\bT(\br)\cdot\beps_0)^2g(\br)d\br},
    \label{fellpticity}
\end{eqnarray}
$\fT(\br,\bu_1,\bu_2)=\bA(\bu_1)\cdot\bT(\br)\cdot\bA(\bu_2)$ and the angular brackets $\indiceGauche{n}{\langle\cdots\rangle}_0$ denote a $n-$body statistical-mechanical average over molecular orientations in the absence of the probing wave. If $W_1(\bu_1)$ and $W_2(\bu_1,\bu_2)$ are the one and two-body orientational probability densities, then
\begin{eqnarray}
\nonumber
    \indiceGauche{1}{\langle\cdots\rangle}_0&=&\int(\cdots)W_{1}(\bu_1)d\bu_1\\
    \nonumber
    \indiceGauche{2}{\langle\cdots\rangle}_0&=&\int(\cdots)W_{2}(\bu_1,\bu_2)d\bu_1d\bu_2
\end{eqnarray}
\end{widetext}
In writing Eq.\eqref{A}, it is tacitly assumed that molecules have symmetric top shape. The orientations of vector $\bu_i$ are generally specified by a pair of spherical polar angles $(\vartheta_i,\varphi_i)$. However, in order to simplify our analysis and since the averages do not involve nor the amplitude electric vector of the incident wave, nor its wavelength, we assume that the orientational probability densities $W_i$ do not depend on $\varphi_i$. This has the particular merit of simplifying the general expressions of $f_{\beps\beps_0}(\kappa)$ without affecting the essential physics. 

In order to proceed, we introduce the three vectors $\bv=(0,0,1)$, $\bh_1=(0,-1,0)$ and $\bh_2=(1,0,0)$ and the four Rayleigh ratios $R_{VV}=R_{\bv\bv}$, $R_{VH}=R_{\bh_1\bv}$, $R_{HV}=R_{\bv\bh_2}=R_{VH}$ and $R_{HH}=R_{\bh_1\bh_2}=R_{\bh_2\bh_1}$. These Rayleigh ratios are relevant for lateral light scattering \cite{BernePecora1970Book}. We also expand $g(\br)$ in Legendre polynomials $P_L(\hat\br)=P_L(\cos\theta_r)$ ($\theta_r$ being the colatitude angle of spherical polar coordinates specifying the orientations of vector $\br$ in the laboratory frame), viz.
\begin{eqnarray}
    g(\br)=\frac{1}{4\pi}\sum_{L=0}^{\infty}G_{L}(r)P_L(\hat{\br}).
    \label{Eq31}
\end{eqnarray}
For further purposes, we also define the parameters
\begin{eqnarray}
    g_L=\int_0^{\infty}\frac{G_L(r)dr}{r^4}.
    \label{Eq32}
\end{eqnarray}
With all these notations in mind, the four Rayleigh ratios pertaining to lateral light scattering may generically be written as follows, 
\begin{eqnarray}
 R_{VV}&=&\frac{\pi^2(\fn^2-1)^2}{\rho_0\lambda_0^4}\left(\fS+\frac{4}{5}\kappa^2g_2^{VV}+\frac{12a}{5}X_3\right)\quad\label{RVVST}\\
    R_{VH}&=&\frac{3\pi^2(\fn^2-1)^2}{5\rho_0\lambda_0^4}\left(\kappa^2g_{2}^{VH}+aX_1\right)=R_{HV}\label{RVHHVST}\\
    R_{HH}&=&\frac{3\pi^2(\fn^2-1)^2}{5\rho_0\lambda_0^4}\left(\kappa^2g_2^{HH}+aX_2\right)\label{RHHST}
\end{eqnarray}
where $a$ is a constant given by
\begin{eqnarray}
    a=\frac{(\fn^2-1)^2}{168\pi^2\rho_0},
    \label{a}
\end{eqnarray}
$X_1$, $X_2$ and $X_3$ are the DID contributions to the Rayleigh ratios \eqref{RVVST}-\eqref{RHHST} that may be written in terms of $g_L$ and $f_{\beps\beps_0}(\kappa)$, viz.
\begin{eqnarray}
     X_1&=&(21g_0+3g_2-4g_4)f_{VH}(\kappa),\label{X1}\\
     X_2&=&(21g_0-6g_2+g_4)f_{HH}(\kappa),\label{X2}\\
    X_3&=&(7g_0+2g_2+2g_4)f_{VV}(\kappa),\label{X3}.
\end{eqnarray}
The ellipticity factors are given by their general (tedious) expressions as given in Appendix \ref{AppA}. We also have introduced the orientational correlation parameters $g_{2}^{VH}$, $g_2^{HH}$ and $g_2^{VV}$ in the manner of Fröhlich (i.e., average of a quantity in the dense phase over average of the same quantity in the ideal gas phase \cite{Frohlich1958Book,Bottcher1973Book}), viz.
\begin{eqnarray}
    g_2^{VH}&=&\frac{15}{2}\indiceGauche{1}{\langle\sin^2\vartheta_1\cos^2\vartheta_1\rangle}_0\label{g2VH}\\
    g_2^{HH}&=&\frac{15}{8}\indiceGauche{1}{\langle\sin^4\vartheta_1\rangle}_0\label{g2HH}
\end{eqnarray}
and
\begin{eqnarray}
\nonumber
   \frac{4}{5}\kappa^2g_2^{VV}=4\kappa^2\indiceGauche{1}{\langle P_2^2(\bu_1)\rangle}_0+4\kappa\;\indiceGauche{1}{\langle P_2(\bu_1)\rangle}\qquad\quad\\+4\kappa(\fS-1)(\;\indiceGauche{2}{\langle P_2(\bu_1)\rangle}_0+\kappa\;\indiceGauche{2}{\langle P_2(\bu_1)P_2(\bu_2)\rangle}_0).\quad
   \label{g2VV}
\end{eqnarray}
By inspection of Eqs.\eqref{g2VH} and \eqref{g2HH}, the correlation parameters $g_2^{VH}$ and $g_{2}^{HH}$ factors are not independent. They are related by
\begin{eqnarray}
    g_{2}^{VH}+4g_{2}^{HH}=5(1-\indiceGauche{1}{\langle P_{2}(\bu_1)\rangle}_0). 
    \label{g2VHHH}
\end{eqnarray}
while $g_2^{VV}$ is a priori an independent correlation parameter. The vertical polarization ratio $r_v$ is given by
\begin{eqnarray}
    r_v&=&\frac{R_{VH}}{R_{VV}}=\frac{3(\kappa^2g_{2}^{VH}+aX_1)}{5\fS'+12aX_3},\label{rvX1X3}
\end{eqnarray}
where $\fS'$ is related to $\fS$ by
\begin{eqnarray}
    \fS'=\fS+\frac{4}{5}\kappa^2g_2^{VV}.
    \label{Sprime}
\end{eqnarray}
The \textit{horizontal} polarization ratio $r_h$ is, by definition \cite{Dezelic1970PureApplChem,Kerker2013Book}
\begin{eqnarray}
    r_h&=&\frac{R_{VH}}{R_{HH}}=\frac{\kappa^2g_{2}^{VH}+aX_1}{\kappa^2g_{2}^{HH}+aX_2}=1\label{rhX1X2}
\end{eqnarray}
since for lateral light scattering, the equality $r_h=1$ is fulfilled independent of the molecular model \cite{Dezelic1970PureApplChem,Kerker2013Book}. 

In order to proceed, we remark that from research group to research group, the parameter which remains more or less the same in the same experimental conditions and given a probing wavelength is the vertical depolarization ratio $r_v$ and the one for unpolarized incident and scattered lights $r_u$. For $90^{\circ}$ light scattering, we have, by the Krishnan relation $r_u=2r_v(1+r_v)^{-1}$ \cite{Kerker2013Book}. Therefore, this suggests to use $r_v$ as an input of the theory (this will free us from computing $r_v$ for $\kappa=0$ when comparing the theory with experimental data). We can then rewrite Eqs.\eqref{rvX1X3} and \eqref{rhX1X2} as the algebraic equations
\begin{eqnarray}
\nonumber
    3a(X_1-4r_vX_3)&=&5r_v\fS'-3\kappa^2g_2^{VH}\\
    a(X_1-X_2)&=&\kappa^{2}(g_2^{HH}-g_2^{VH})
    \label{X1X2X3}
\end{eqnarray}
This system of equations, by inspection of Eqs.\eqref{X1}-\eqref{X3}, can be interpreted as a system of two linear algebraic equations the unknowns of which are $g_2$ and $g_4$. Therefore, $g_2$ and $g_4$ can always be determined in terms of $g_0$, $r_v$, $\fS$, $\kappa$, the orientational correlation parameters $g_{2}^{VV}$, $g_{2}^{VH}$, $g_{2}^{HH}$ and extra two-body orientational averages. It is therefore, despite its "simplicity", a mathematical problem which has a cumbersome explicit solution, where one has to compute one- and two-body orientational averages. This is not, as may be seen, for example, from our recent work \cite{Dejardin2022PRE1} an easy task to achieve as we have to find a general method of computation of all unknown parameters, together with a potential describing intermolecular interactions (of medium to short range) as correctly as is possible. However, for our present purposes, we do not need such a solution, as we shall see soon. This is particularly so when $\kappa=0$ or when DID is ignored, where an \textit{exact} solution can be provided. Before proceeding, we introduce the DID-induced relative polarizability anisotropy given by
\begin{eqnarray}
    \kappa_{DID}^2=21ag_0=\frac{(\fn^2-1)^2g_0}{8\pi^2\rho_0}=\frac{(\fn^2-1)^2\Delta\alpha_{DID}^2}{9\bar\alpha^2}.\quad
    \label{kappaDID}
\end{eqnarray}
This parameter is similar in form with Eq.(50) of Kielich's paper \cite{Kielich1967JCP}.
\section{Rayleigh ratios for dense isotropically polarizable molecules}\label{3}
As just mentioned, this is a situation that may be solved exactly \textit{because} $\kappa=0$, entailing that the ellipticity factors are unity (see Appendix \ref{AppA}), that $\fS'=\fS$ and that the orientational correlation factors are irrelevant so that they all can be taken equal to unity without affecting the final result. The system of Eqs.\eqref{X1X2X3} yield $X_1=X_2$ and $3a(X_1-4r_vX_3)=5r_v\fS$, yielding for $g_2$ and $g_4$ the expressions
\begin{eqnarray}
    g_2&=&\frac{5\kappa_{DID}^2(3-4r_v)-25r_v\fS}{21a(3+16r_v)}\quad\label{g2Kappa0}\\
    g_4&=&\frac{3\kappa_{DID}^2(3-4r_v)-15r_v\fS}{7a(3+16r_v)}\quad
    \label{g4Kappa0}
\end{eqnarray}
Particularly, if the structure of the liquid is spherically symmetric, $g_2=g_4=0$ because $g(\br)$ does not depend on $\br$ orientations. In this quite special situation, we have
\begin{eqnarray}
\nonumber
    r_v=\frac{3\kappa_{DID}^2}{5\fS+4\kappa_{DID}^2}
\end{eqnarray}
However, this is generally not the case. For example, it was shown by X-ray and neutron scattering that liquid CCl$_4$ has, in reality, a complicated structure \cite{Narten1976JCP},  strongly suggesting that $g(\br)$ in effect \textit{depends} on $\br$ orientations. Therefore, $g_2$ and $g_4$ \textit{cannot} be taken equal to zero despite they may be found small in regard with $g_0$. Hence, $g_2\neq0$ and $g_4\neq0$ and we have 
\begin{eqnarray}
    r_v=\frac{3\kappa_{DID}^2 (1-x)}{5\fS+4\kappa_{DID}^2(1+4x)},\;x=\frac{g_2}{5g_0}.
    \label{rvFullDID}
\end{eqnarray}
However, this equation is of little use here since $r_v$ is generally deduced from experimental measurements. We can then combine Eqs.\eqref{RVVST}-\eqref{RHHST} with Eqs.\eqref{X1}-\eqref{X3} with $\kappa=0$ to obtain $R_{VV}$ and $R_{VH}$ as
\begin{eqnarray}
    R_{VV}&=&\frac{3\pi^2}{\rho_0\lambda_0^4}\frac{(\fn^2-1)^2(4\kappa_{DID}^2+\fS)}{3+16r_v}\label{RVVDID}\\
    \nonumber
    R_{VH}&=&\frac{3\pi^2}{\rho_0\lambda_0^4}\frac{(\fn^2-1)^2r_v(4\kappa_{DID}^2+\fS)}{3+16r_v}\\&=&R_{HV}=R_{HH}.\label{RVHHVDID}
\end{eqnarray}
These equations are of remarkable simplicity and entail that
\begin{eqnarray}
    \nonumber
     3R_{VV}+16R_{VH}=R_0
\end{eqnarray}
where $R_0$ is entirely independent of $r_v$ (therefore of $g_2$). We have
\begin{eqnarray}
    R_0
    =\frac{3\pi^2}{\rho_0\lambda_0^4}(\fn^2-1)^2(4\kappa_{DID}^2+\fS).
    \label{R0eq}
\end{eqnarray}
This very simple equation can be used in order to obtain an experimental determination of $\kappa_{DID}$ when $\kappa=0$. Eqs.\eqref{RVVDID}-\eqref{R0eq} are amongst the central results of this work. $\fL^4$ scaling of the Rayleigh ratios is obtained only if $4\kappa_{DID}^2\gg\fS$. In the opposite situation, $\fL^2$ scaling obtains, while in the general case (and this is so for CCl$_4$ as we show later), \textit{no trivial scaling can be found}. 

\section{Rayleigh ratios for dense anisotropically polarizable symmetric top molecules with zero DID}\label{4}
When, ideally, the DID components are zero (but we insist that this is \textit{never} so), then Eqs. \eqref{X1X2X3} imply that $g_2^{VH}=g_2^{HH}$. Moreover, with the intention of checking the theory in the range $293-298$ K, one may use the (Vlasov) rotational mean field approximation, viz. $W_2(\vartheta_1,\vartheta_2)\approx W_1(\vartheta_1)W_1(\vartheta_2)$. Therefore, in this context we have
\begin{eqnarray}
    g_2^{VH}&=&1-\indiceGauche{1}{\langle P_2(\bu_1)\rangle}_0\label{g2VHRot}\\
    \fS'&=&\fS[1+2\kappa(1-g_2^{VH})]^2+\frac{4}{5}\kappa^2g_{2}^{VH}(6-5g_2^{VH})\qquad\label{fSPrimeRot}
\end{eqnarray}
so that only \textit{one} orientational correlation parameter is required in order to describe all Rayleigh ratios. The vertical depolarization ratio is
\begin{eqnarray}
    r_v=\frac{3\kappa^2g_{2}^{VH}}{5\fS\left(1+2\kappa(1-g_{2}^{VH})\right)^2+4\kappa^2g_{2}^{VH}(6-5g_{2}^{VH})}.\qquad
    \label{rvRotZeroDID}
\end{eqnarray}
Here, Eq.\eqref{rvRotZeroDID} may be used with advantage if $\kappa$ is known from quantum ab-initio calculations or from Tables and if $\chi_T$ is also known from independent measurements. Since experimental data provide $r_v$, this equation is a quadratic one for $g_2^{VH}$ having, by inspection of its coefficients, two roots of opposite signs. Since $0<g_{2}^{VH}\leq 1$ due to the upper bound of the average second Legendre polynomial (which is also $1$), it is straightforward to calculate the adequate root. Importantly, this avoids computing $g_2^{VH}$ via complicated integrals. The correlation parameter $G$ is related to $g_{2}^{VH}$ via \cite{Dezelic1970PureApplChem}
\begin{eqnarray}
\nonumber
    \frac{\kappa^2G}{\fS}=\frac{5r_v}{3-4r_v}\qquad\qquad\qquad\\
    =\frac{\kappa^2g_2^{VH}}{\fS[1+2\kappa(1-g_2^{VH})]^2+4\kappa^2g_{2}^{VH}(1-g_{2}^{VH})}
    \label{Grot}
\end{eqnarray}
where Eq.\eqref{rvRotZeroDID} has been used in order to establish the second equality. This last equation shows that $G=1$ when $g_2^{VH}=1$. The Rayleigh ratios $R_{VV}^{Rot}$ and $R_{VH}^{Rot}$ are then given by
\begin{eqnarray}
    R_{VV}^{Rot}=\frac{\pi^2(\fn^2-1)^2}{\rho_0\lambda_0^{4}}\fS',\qquad
    \label{RVVRotZeroDID}
\end{eqnarray}
where $\fS'$ is given by Eq.\eqref{fSPrimeRot} and
\begin{eqnarray}
R_{VH}^{Rot}=R_{HH}^{Rot}&=&\frac{3\pi^2(\fn^2-1)^2\kappa^2g_{2}^{VH}}{5\rho_0\lambda_0^{4}}
    \label{RVHRotZeroDID}
\end{eqnarray}
Both Eqs.\eqref{RVVRotZeroDID} and \eqref{RVHRotZeroDID} reduce to the well-known Gans results \cite{Gans1921AnnPhys} for $\fS\rightarrow1$ and $g_{2}^{VH}=1$, and to the Einstein ones (divided by $\fL_L^2$ due to his scaling of scattered intensities by the squared Maxwell field \cite{Einstein1910AnnPhys,Pabst2024JCP}) for $\kappa=0$, as in this situation, $\rho_0\lambda_{0}^4R_{VV}=\pi^2(\fn^2-1)^2\fS$ and $R_{VH}=0$. Here, ultimately, the rotational $R_{VV}^{Rot}$ \eqref{RVVRotZeroDID} shows that when $R_{VH}^{Rot}$ is written as given by Eq.\eqref{RVHRotZeroDID}, its writing is not as trivial when orientational correlations are present. Likewise, $r_v$ given by Eq.\eqref{rvRotZeroDID} is nontrivial. Nevertheless, it can be reduced to Eq.\eqref{rvG} if Eq.\eqref{Grot} specifying the relation between $g_{2}^{VH}$ and $G$ is used. At last, we find that rotational contributions to the Rayleigh ratios \textit{scale as} $\fL^2$ and not as $\fL^4$, as is usual in statistical-mechanical formulations of the problem. Nevertheless, as DID is not involved in Eqs.\eqref{RVVRotZeroDID} and seq., conclusions regarding scaling of Rayleigh ratios with $\fL$ are \textit{premature}. Now, we are are ready to handle the situation where both rotation and DID mechanisms mix. 

\section{Mixed situations}\label{5}
As already mentioned, the general solution of Eqs.\eqref{X1X2X3} is always possible, but certainly cumbersome. Furthermore, it may be of no use for further practical purposes. Therefore, we content ourselves with deriving rotational corrections to Eqs.\eqref{RVVDID} and \eqref{RVHHVDID} on one hand (rotation-corrected DID), and DID corrections to Eqs.\eqref{RVVRotZeroDID} and \eqref{RVHRotZeroDID}  (DID-corrected rotation). 
\subsection{Rotation-corrected DID formulas}
In this situation, the light scattering process is almost entirely governed by DID, therefore we can set $g_2^{VH}=g_{2}^{VV}=g_2^{HH}\approx1$ and neglect their $\kappa$ dependence. Therefore Eqs.\eqref{X1X2X3} yield $X_1=X_2$ and $3a(X_1-4r_vX_3)=5r_v\fS-3\kappa^2$. The full expressions for the ellipticity factors $f_{\beps\beps_0}(\kappa)$ must be used, with $W_1=(4\pi)^{-1}$, $W_2=(4\pi)^{-2}$, giving rise to cumbersome expressions for $X_1,X_2$ and $X_3$. The expressions are not provided, however, solving again for $g_2$ and $g_4$, construct $X_1$, $X_2$ and $X_3$ and expand to order $\kappa^2$ leads after tedious algebra to simple expressions for $R_{VV}$ and $R_{VH}$, viz.
\begin{eqnarray}
\nonumber
    R_{VV}&=&\frac{3\pi^2}{\rho_0\lambda_0^4}\frac{(\fn^2-1)^2}{3+16r_v}\left(4\kappa_{DID}^2+\fS+\frac{48\kappa^2}{5(3+16r_v)}\right)\label{RVVRotCorrDID}\\
    \\
    \nonumber
    R_{VH}&=&\frac{3\pi^2}{\rho_0\lambda_0^4}\frac{(\fn^2-1)^2r_v}{3+16r_v}\left(4\kappa_{DID}^2+\fS+\frac{48\kappa^2}{5(3+16r_v)}\right)\\&=&R_{HV}=R_{HH}.\label{RVHHVRotCorrDID}
\end{eqnarray}
Higher corrections are $O(\kappa^4)$ and can therefore be neglected. These equations are valid if
\begin{eqnarray}
    \kappa^2<\frac{5(4\kappa_{DID}^2+\fS)(3+16r_v)}{48}
    \label{RvalRotCorrDID}
\end{eqnarray}
which, having determined $\kappa_{DID}^2$ from experimental data, is easy to check by self-consistency. An expression for $r_v$ itself is of little use as we consider that $r_v$ is an input when DID is the main mechanism of light scattering. In this situation, $r_v$ is approximately given by Eq.\eqref{rvFullDID}, and improvement on this equation is not required here. To conclude with this subsection, the Rayleigh ratios do not generally scale as $\fL^4$, nor they scale as $\fL^2$. They nevertheless scale as $\fL^4$ if the term involving $\kappa_{DID}^2$ is much larger than all the others. We now treat the situation where the DID contribution acts as a correction to the rotational one.

\subsection{DID-corrected rotation}
This situation is by far the most difficult one as it consists of obtaining an asymptotic expansion to the system of Eqs.\eqref{X1X2X3}. The difficulty here is that the small parameter is not readily identified. Nevertheless, the relevant quantities \textit{are even} in $\kappa_{DID}$, because $X_i$ can generally be written as a linear combination of $g_0$, $g_2$ and $g_4$. That is, it involves $\kappa_{DID}^2$ only and its higher positive integer powers. Here we use the following trick. The asymptotic expansion that we seek is based upon speculating that one must have, in particular, $aX_1\ll\kappa^2g_2^{VH}$ and $12aX_3\ll5\fS'$, quite special conditions which however leave $r_v$ given by Eq.\eqref{rvRotZeroDID} \textit{unaffected}. As we shall see later, $r_v$ is also unaffected in the Dezelic-Vavra semi-empirical work \cite{DezelicVavra1966CroatChemActa} which allows the correct value of $\kappa$ to be extracted from dilution experiments. Then, we must have nonzero $X_1$, $X_2$ and $X_3$ such that $r_v$ is \textit{approximately not altered by DID}. In other words, we must have
\begin{eqnarray}
    \nonumber
    r_v=\frac{3(\kappa^2g_{2}^{VH}+aX_1)}{5\fS'+12aX_3}=\frac{3\kappa^2g_2^{VH}}{5\fS'}
\end{eqnarray}
which translates into the first sought equation, viz.
\begin{eqnarray}
    X_1=4r_vX_3
    \label{X1X3DIDCorrRot}
\end{eqnarray}
where in effect, $r_v=3\kappa^2g_2^{VH}/(5\fS')$ (this can be checked by insertion). Since the same must hold for $r_h$ (which, in reality, is never altered since for lateral light scattering, it is always $1$), we have
\begin{eqnarray}
    \nonumber
    r_h=\frac{\kappa^2g_{2}^{VH}+aX_1}{\kappa^2g_{2}^{HH}+aX_2}=\frac{g_2^{VH}}{g_2^{HH}}=1
\end{eqnarray}
which implies that $g_2^{VH}=g_{2}^{HH}$, therefore again that 
\begin{eqnarray}
    X_1=X_2.
    \label{X1X2DIDCorrRot}
\end{eqnarray}
This equation completes Eq.\eqref{X1X3DIDCorrRot}. Because $r_v$ given by Eq.\eqref{rvRotZeroDID}, it is therefore not affected and maintains its expression if we maintain the rotational mean field approximation, which we assume to be valid from now on. The determination of $g_2$, $g_4$ and $X_i$ proceed in the same way as in the preceding Sections. Moreover, at the correction order we consider here, the DID contribution is approximately isotropic, hence, we essentially have
\begin{widetext}
\begin{eqnarray}
    f_{VV}(\kappa)=f_{VH}(\kappa)=f_{HH}(\kappa)\approx 1
    \label{FRotCorrDID}
\end{eqnarray}
Thus, solving Eqs.\eqref{X1X3DIDCorrRot} and \eqref{X1X2DIDCorrRot} we obtain the simple expressions for $g_2$ and $g_4$, namely
\begin{eqnarray}
    g_2&=&\frac{5\kappa_{DID}^2(3-4r_v)}{21a(3+16r_v)}\label{g2DIDCorrRot}\\
    g_4&=&\frac{3\kappa_{DID}^2(3-4r_v)}{7a(3+16r_v)}\label{g4DIDCorrRot}
\end{eqnarray}
leading, using Eqs.\eqref{RVVST}-\eqref{g2VV} of Section \ref{3}, Eqs. \eqref{fSPrimeRot}, the rotational mean field approximation $W_2(\vartheta_1,\vartheta_2)\approx W_1(\vartheta_1)W_1(\vartheta_2)$ and the ellipticity factors \eqref{FRotCorrDID}, to the following expressions for the DID-corrected rotation expressions for $R_{VV}$ and $R_{VH}$, viz.
\begin{eqnarray}
    R_{VV}&=&\frac{\pi^2(\fn^2-1)^2}{\rho_0\lambda_0^{4}}\fS'\left\{1+\frac{4\kappa_{DID}^2}{\fS\left[1+2\kappa(1-g_2^{VH})\right]^2+4\kappa^2g_2^{VH}\left(2-g_2^{VH}\right)}+\cdots\right\}\label{RVVDIDCorrRot}\\
    R_{VH}&=&\frac{3\pi^2(\fn^2-1)^2\kappa^2g_{2}^{VH}}{5\rho_0\lambda_0^{4}}\left\{1+\frac{4\kappa_{DID}^2}{\fS\left[1+2\kappa(1-g_2^{VH})\right]^2+4\kappa^2g_2^{VH}\left(2-g_2^{VH}\right)}+\cdots\right\}\label{RVHDIDCorrRot}
\end{eqnarray}
\end{widetext}
$r_v$ being still given by Eq.\eqref{rvRotZeroDID}. The results derived in this subsection are valid if
\begin{eqnarray}
\nonumber
    4\kappa_{DID}^2&<&\fS\left[1+2\kappa(1-g_2^{VH})\right]^2\\&+&4\kappa^2g_2^{VH}\left(2-g_2^{VH}\right).
    \label{RangeValidityDIDCorrRot}
\end{eqnarray}
Since, when $\kappa$ is well-known, $R_{VV}^{Rot}$ and $R_{VH}^{Rot}$ are not very difficult to evaluate from the experimental value of $r_v$ as explained in Section \ref{4}, we may expect to establish a simple criterion in order to decide whether the DID or rotational component dominates light scattering. In effect, Eq.\eqref{RangeValidityDIDCorrRot} with Eq.\eqref{kappaDID} show that Eqs.\eqref{RVVDIDCorrRot} and \eqref{RVHDIDCorrRot} apply if
\begin{eqnarray}
\nonumber
    \psi<1,\quad\text{and}\quad\Psi<1
\end{eqnarray}
where
\begin{eqnarray}
    \psi&=&\frac{5r_v(\fn^2-1)^2}{16\pi\kappa^2(3+16r_v)}\label{psi}\\
    \Psi&=&\frac{R_{VV}^{Exp}}{R_{VV}^{Rot}}-1\label{Psi}
\end{eqnarray}
where $R_{VV}^{Exp}$ is the value of $R_{VV}$ deduced from experiment and $R_{VV}^{Rot}$ is computed from Eq.\eqref{RVVRotZeroDID} as indicated before. We note in passing that $\psi<1$ is similar to an earlier criterion proposed by Cummins et al. on the basis of numerical simulations of Pick and co-workers \cite{Cummins1996PRE,Pick1994JCP}. Here, the criteria Eqs.\eqref{psi} and, particularly \eqref{Psi}, are directly related to light scattering experimental data. Importantly, these criteria do not require the specification of any intermolecular interaction potential.
\section{Comparison with experimental data}\label{6}
Because our results are entirely analytical, we find worth testing them directly against available experimental data on some liquids. Another reason for proceeding so is that $G_0(r)$ is \textit{not} equal to $4\pi$ times the center-to-center radial pair distribution function $g_r(r)$ as it would seem at first glance by overlooking Eq.\eqref{Eq31}. This relation is furthermore not so easy to establish.

Amongst the plethora of liquids which have been characterized via Rayleigh light scattering, there are several which are archetypal, in which we are interested here. These are carbon tetrachloride CCl$_4$ which is typical of a nonpolar liquid and made of isotropically polarizable molecules, of Benzene C$_6$H$_6$ and Toluene C$_7$H$_8$ because they are more than frequently used as reference liquids for the light scattering experiments, and Carbon disulfide CS$_2$ and Nitrobenzene C$_6$H$_5$NO$_2$ which are very good scatterers, and as we shall see are quite representative of both situations described in Section \ref{5}. The properties of the liquids we consider are summarized in Table \ref{Table1}. The $\kappa$ values are those for the single molecule (in the ideal gas phase) and are taken from Tables \cite{Timmermans}. We have checked that those values do not differ much from quantum ab-initio calculations \cite{PabstData2024} ($2\%$ difference in absolute value).

\begin{table}[h!]
   \setlength{\tabcolsep}{0.1cm}
   \centering
   \begin{tabular}{ccccccccccccc}
    \hline
    \hline
Liquid & $\rho_0\times10^{-21}$cm$^{-3}$ & $\chi_T\times10^{12}$cm$^2$dyn$^{-1}$ & $\kappa^2$ \\
\hline  
CCl$_4$              & 6.205 & 108  &  0   \\
\hline
C$_6$H$_6$           & 6.736 & 97 & 0.0324   \\
\hline
C$_7$H$_8$           & 5.636 &  92  & 0.0402   \\
\hline
CS$_2$               & 9.935 &  96  & 0.134   \\
\hline
C$_6$H$_5$NO$_2$     & 5.863 &  50  & 0.0502  \\
\hline
    \end{tabular}
    \caption{Number density, isothermal compressibility of some liquids together with the squared relative polarizability of their individual constitutive molecules. $\kappa$ values calculated from Tables \cite{Timmermans}.}
    \label{Table1}
  \end{table} 

In our comparison with experiment, we use refractive index values provided by Chang et al.\cite{Chang2024OptExpress} which are valid for $T=295$ K, and neglect their temperature variation. This is reasonable as we restrict to temperatures between $293$ K and $298$ K. Because of this restriction, we also neglect the temperature variation of $\chi_T$. Next, light sources utilized in order to perform light scattering experiments are either green or red-colored. Therefore, there is not really use to compare our theoretical outcomes for wavelengths corresponding to blue light, where the results for the Rayleigh ratios vary from group to group. Hence, in our comparison, we do not systematically consider experimental data corresponding to wavelengths lying below $500$ nm, except otherwise stated. Then we focus on $R_{VV}$ experimental values, which can be constructed from the model-independent equation
\begin{eqnarray}
    R_{VV}^{Exp}=\frac{6R_{iso}}{3-4r_v}.
    \label{RVVRisoExp}
\end{eqnarray}
A study of Rayleigh ratios $R_u$ corresponding to unpolarized incident and scattered lights together with the corresponding depolarization ratios $r_u$ at $\lambda_0=633$ nm and $T=295$ K has been performed by Pike et al.\cite{Pike1975JCP}, for all liquids listed in Table \ref{Table1} save Nitrobenzene. $R_{iso}$, and $R_u$ are related by the Cabannes equation, viz.
\begin{eqnarray}
    R_{iso}=\frac{6-7r_u}{6+6r_u}R_u
    \label{Cabannes}
\end{eqnarray}
while $r_u$ and $r_v$ are related, as already mentioned by Krishnan's relation (here specialized to $90^{\circ}$ light scattering) \cite{Kerker2013Book}, viz.
\begin{eqnarray}
    r_u=\frac{2}{1+r_v^{-1}}
    \label{Krishnan}
\end{eqnarray}
so that the experimental value $R_{VV}^{Exp}$ to which we compare our theoretical findings can be computed from Eq.\eqref{RVVRisoExp}. This means, in proceeding so, that we \textit{assume}, rightly or wrongly, that $R_u$ values are free of ill-controlled experimental errors. Light scattering data for all these liquids are indicated in Table \ref{Table2}.
\begin{widetext}

\begin{table}[h!]
   \setlength{\tabcolsep}{0.4cm}
   \centering
   \begin{tabular}{ccccccccccccc}
    \hline
    \hline
Liquid & $\lambda_0$ (nm)& $T$ (K)& $r_u$ & $R_u$ & $R_{iso}$ & Reference & $\fn$ & $r_v$ & $R_{VV}^{Exp}$ \\
\hline  
CCl$_4$              &546&298& 0.049& 5.53& 4.97&\citenum{Dezelic1970PureApplChem}&1.460 & 0.025 & 10.3  \\
                     &633&295& 0.042& 2.95& 2.69&\citenum{Pike1975JCP}&1.456&0.021&5.54\\
\hline
C$_6$H$_6$           &546&296&0.42 & 15.8&5.67&\citenum{Coumou1964TransFarSoc}&1.503&0.266&17.6\\
                     &633&295&0.432&8.51&2.94&\citenum{Pike1975JCP}&1.496&0.276&9.32\\
                     &694&295&0.439&5.86&1.99&\citenum{Pike1975JCP}&1.493&0.281&6.36\\
\hline
C$_7$H$_8$           &546&298&0.48&17.6&5.23& \citenum{Dezelic1970PureApplChem}&1.499&0.316&18.1   \\
                     &633&295&0.528&10.4&2.61&\citenum{Pike1975JCP}&1.492&0.359&10.0\\
                     &532&298& $\cdots$ & $\cdots$ & $\cdots$ &\citenum{Takahashi2019ApplSci}  &1.500&0.316&23.4\\
\hline
CS$_2$               &546&298&0.656&84.6&11.98& \citenum{Dezelic1970PureApplChem}&1.631&0.488&68.66\\
                     &633&295&0.65&42.6&6.24&\citenum{Pike1975JCP}&1.617&0.481&34.85\\
\hline
C$_6$H$_5$NO$_2$     &546&298&0.74&65.8&5.17&\citenum{Dezelic1970PureApplChem}&1.557&0.587&47.64\\
\hline
    \end{tabular}
    \caption{Light scattering data for all liquids considered in this work. Rayleigh ratios in units of $10^{-6}$ cm$^{-1}$. All refractive indices from Chang et al.\cite{Chang2024OptExpress}.}
    \label{Table2}
  \end{table} 
  
In addition to Dezelic's and Pike's data for Toluene, we have also provided $R_{VV}^{Exp}$ for $\lambda_0=532$ nm at $298$ K as determined by Takahashi et al. \cite{Takahashi2019ApplSci}. This value was \textit{only recently} experimentally determined \textit{with extreme care} by these authors. As we shall see soon, this value is very useful. Now, following Pike et al. \cite{Pike1975JCP} the variation of $r_v$ with wavelength is rather unlikely at a given temperature. This is why we will only retain an average value of $g_{2}^{VH}$ as well as an average value of $\Delta\alpha_{DID}$ as defined by Eq.\eqref{kappaDID}. To determine these averages, we first must know which formula for $R_{VV}$ given in Section \ref{5} we should apply. Therefore, we begin by determining $R_{VV}^{Rot}$ from Eq.\eqref{RVVRotZeroDID}, which requires solving the quadratic equation \eqref{rvRotZeroDID} for each $r_v$ value given in Table \ref{Table2}, given the liquid at study, without worrying about the precise value obtained for $g_2^{VH}$ for the moment. This gives us in turn values of $\vert\Psi\vert$ and $\psi$, which will give us in turn further information regarding which, Eq.\eqref{RVVRotCorrDID} or Eq. \eqref{RVVDIDCorrRot} we should use in order to extract $\Delta\alpha_{DID}$ from $R_{VV}^{Exp}$. The results of these computations are given in Table \ref{Table3}.

\begin{table}[h!]
   \setlength{\tabcolsep}{1cm}
   \centering
   \begin{tabular}{ccccccccccccc}
    \hline
    \hline
Liquid & $\lambda_0$ (nm)& $T$ (K)& $R_{VV}^{Rot}$ & $\vert\Psi\vert$ & $\psi$ \\
\hline  
CCl$_4$              &546&298&6.31&0.63&$\infty$\\
                     &633&295&3.39&0.63&$\infty$ \\
\hline
C$_6$H$_6$           &546&296&17.2&0.03&0.18\\
                     &633&295&8.87&0.05&0.18\\
                     &694&295&5.98&0.06&0.17\\
\hline
C$_7$H$_8$           &532&298&23.2&0.01&0.15\\
                     &546&298&20.8&0.09&0.15\\
                     &633&298&10.3&0.02&0.15\\
\hline
CS$_2$               &546&298&49.3&0.39&0.09\\
                     &633&295&26.1&0.34&0.09\\
\hline
C$_6$H$_5$NO$_2$     &546&298&19.8&1.41&0.2\\
\hline
    \end{tabular}
    \caption{$R_{VV}^{Rot}$ from Eq.\eqref{RVVRotZeroDID} in units of $10^{-6}$ cm$^{-1}$ together with $\vert\Psi\vert$ and $\psi$ values. Surprisingly, for Nitrobenzene $\Psi=141\%$ despite a well-defined $\kappa$ (see text).}
    \label{Table3}
  \end{table} 
    
As it appears from Table \ref{Table3}, Eq.\eqref{RVVDIDCorrRot} cannot be applied for CCl$_4$ which was expected since light scattering arises solely from DID in this case. However, more surprisingly, this equation \textit{does not apply to Nitrobenzene either}, since $\vert\Psi\vert=141\%$ despite the Nitrobenzene molecule \textit{has a well-defined, non-zero} $\kappa$. It follows that both for CCl$_4$ and C$_6$H$_5$NO$_2$, $\Delta\alpha_{DID}$ must be determined from the equation 
\begin{eqnarray}
    \frac{\Delta\alpha_{DID}}{3\bar\alpha}=\frac{1}{(\fn^2-1)}\sqrt{\frac{\rho_0\lambda_0^4(3+16r_v)R_{VV}^{Exp}}{12\pi^2(\fn^2-1)^2}-\frac{\fS}{4}-\frac{12\kappa^2}{5(3+16r_v)}}\qquad
    \label{DeltaAlphaDIDdet1}
\end{eqnarray}
which is the sole possible remaining formula at our disposal. For the three other liquids, $\Delta\alpha_{DID}$ is estimated from $R_{VV}^{Exp}$ via

\begin{eqnarray}
    \frac{\Delta\alpha_{DID}}{3\bar\alpha}=
    \frac{\sqrt{\Psi\left[\fS(1+2\kappa(1-g_{2}^{VH}))^2+4\kappa^{2}g_{2}^{VH}(2-g_{2}^{VH})\right]}}{2(\fn^2-1)},\qquad
    \label{DeltaAlphaDIDdet2}
\end{eqnarray}
again without worrying about the (small, less than $2\%$ in absolute value across $100$ nm wavelength variation) wavelength dependence of $g_{2}^{VH}$ extracted from Eq.\eqref{rvRotZeroDID} interpreted as a quadratic equation for $g_2^{VH}$. Then, in both situations, having collected all the possible $\Delta\alpha_{DID}$ values for a given substance, we retain an algebraic average $\overline{\Delta\alpha_{DID}}$. We also proceed the same way for $g_2^{VH}$, retaining only an algebraic average $\overline{g_2^{VH}}$. The values $\overline{\Delta\alpha_{DID}}$ and $\overline{g_2^{VH}}$ give rise to an average value $\overline{R_{VV}}$ which is to be compared to $R_{VV}^{Exp}$ and a depolarization ratio $\overline{r_v}$ from which we define an average value $\overline{R'_{VH}}=\overline{r_v}\times\overline{R_{VV}}$. Explicitly, for the DID-corrected rotation formulas, we have
\begin{eqnarray}
\nonumber
    \overline{R_{VV}}&=&\frac{\pi^2(\fn^2-1)^2}{5\rho_0\lambda_0^4}\left[5\fS(1+2\kappa(1-\overline{g_2^{VH}}))^2+4\kappa^2\overline{g_2^{VH}}(6-5\overline{g_2^{VH}})\right]\\
    &\times&\left\{1+\frac{4(\fn^2-1)^2(\overline{\Delta\alpha_{DID}})^2}{9\bar\alpha^2[\fS(1+2\kappa(1-\overline{g_2^{VH}}))^2+4\kappa^2\overline{g_2^{VH}}(2-\overline{g_2^{VH}})]}\right\}\label{TheorCompExpRVV}\\
    \overline{r_{v}}&=&\frac{3\kappa^2\overline{g_2^{VH}}}{5\fS\left(1+2\kappa(1-\overline{g_2^{VH}})\right)^2+4\kappa^2\overline{g_2^{VH}}(6-5\overline{g_2^{VH}})}\label{Theorrv}\\
    \overline{R_{VH}}&=&\frac{3\pi^2(\fn^2-1)^2\kappa^2\overline{g_2^{VH}}}{5\rho_0\lambda_0^4}\left\{1+\frac{4(\fn^2-1)^2(\overline{\Delta\alpha_{DID}})^2}{9\bar\alpha^2[\fS(1+2\kappa(1-\overline{g_2^{VH}}))^2+4\kappa^2\overline{g_2^{VH}}(2-\overline{g_2^{VH}})]}\right\}\label{TheorCompExpRVH}
\end{eqnarray}
Of course, $\overline{R_{VH}}$ should not deviate too much from $ \overline{R'_{VH}}=\overline{r_v}\times \overline{R_{VV}}$ in order for all this to have a meaning at all (see Table \ref{Table4}). 
\noindent
The rotation-corrected DID formulas that are used are
\begin{eqnarray}
\nonumber
    \overline{R_{VV}}&=&\frac{3\pi^2}{\rho_0\lambda_0^4}\frac{(\fn^2-1)^2}{3+16\overline{r_v}}\left(\frac{4(\fn^2-1)^2}{9\bar\alpha^2}(\overline{\Delta\alpha_{DID}})^2+\fS+\frac{48\kappa^2}{5(3+16\overline{r_v})}\right)\label{RVVRotCorrDIDExp}\\
    \\
    \nonumber
    R_{VH}&=&\overline{r_v}\times \overline{R_{VV}}\label{RVHHVRotCorrDIDExp}
\end{eqnarray}
where in these last two equations $ \overline{r_{v}}$ is the algebraic average of the $r_v$ values for a given liquid that we have at our disposal. We apply these last two equations to liquid CCl$_4$ and C$_6$H$_5$NO$_2$. The numerical details involved are quite tedious, therefore they are not given here. We show in Table \ref{Table4} the results of our computations, and repeat $R_{VV}^{Exp}$ values given in Table \ref{Table2} for ease of reading. 

\begin{table}[h!]
   \setlength{\tabcolsep}{0.3cm}
   \centering
   \begin{tabular}{ccccccccccccc}
    \hline
    \hline
Liquid & $\overline{g_2^{VH}}$& $\bar{\alpha}(\text{\AA}^3)$& $\overline{\Delta\alpha_{DID}}(\text{\AA}^3)$ &  $\lambda_0$ (nm)& $T$ (K)& $\overline{r_v}$ & $\overline{R_{VV}}$& $R_{VV}^{Exp}$ & $\overline{R_{VH}}$& $\overline{R'_{VH}}$\\
\hline  
CCl$_4$              &1  &10.50&2.1&546&298&0.023&10.27&10.3&0.236&0.236 \\
                     & &    &   &633&295&     &5.48 &5.54&0.126&0.126 \\
\hline
C$_6$H$_6$           &0.9 &10.32&1.08&546&296&0.274&17.43&17.6&4.79&4.79\\
                     &    &     &    &633&295&0.275&9.30 &9.32&2.56&2.56\\
                     &    &     &    &694&295&0.275&6.34 &6.36&1.74&1.74\\
\hline
C$_7$H$_8$           &0.9 &12.26&0.63&532&298&0.327&22.94&23.4&7.49&7.49\\
\hline
CS$_2$               &0.97&8.74 &3.67&546&298&0.486&68.01&68.66&33.04&33.04\\
                     &    &     &    &633&295&0.487&35.00&34.85&17.05&17.05\\
\hline
C$_6$H$_5$NO$_2$     &1   &12.92&9.24&546&298&0.587&47.6&47.6&28.0&28.0\\
\hline
    \end{tabular}
    \caption{Average Rayleigh ratio $\overline{R_{VV}}$ in units of $10^{-6}$ cm$^{-1}$ compared with $R_{VV}^{Exp}$. The quantities $\overline{R_{VH}}$ and $\overline{R'_{VH}}$ are strictly equal in value.}
    \label{Table4}
  \end{table} 

\noindent
From Table \ref{Table4}, we remark that when different from $1$, $g_{2}^{VH}<1$, which suggests the reason why in the liquid phase $\kappa_{\ell}^2=\kappa^2g_{2}^{VH}$ is \textit{always} smaller than the gas phase value $\kappa^2$ (and if not, molecular rotation does not dominate the light scattering phenomenon). Moreover, in Nitrobenzene we have $\Delta\alpha_{DID}/\Delta\alpha\approx1.064$ (from Tables \cite{Timmermans}, $\Delta\alpha$ can be evaluated), and since here, the DID mechanism dominates rotation, disagreement with the quantum ab-initio calculation of $\Delta\alpha$ is not surprising : in this substance, light scattering just hardly tracks molecular rotational motion, but rather, DID is responsible for light scattering. In addition, the quantity
\begin{eqnarray}
\nonumber
    3\bar{\alpha}\sqrt{\frac{5r_v\fS}{3-4r_v}}\approx9.02\;\text{\AA}^3
\end{eqnarray}
agrees with our value of $\overline{\Delta\alpha_{DID}}$ for this liquid. However, it exceeds the natural molecular polarizability anisotropy of Nitrobenzene, $\Delta\alpha\approx8.68\;\text{\AA}^3$. Therefore, the ab-initio calculation is, of course, not the cause of the discrepancy \cite{Pabst2022JCP}. 

We have not included in our determination of $\overline{\Delta\alpha_{DID}}$ for C$_7$H$_8$ the values of $R_{VV}^{Exp}$ earlier found for $\lambda_0=546$ nm and $\lambda_0=633$ nm, because they are smaller than the calculated values $R_{VV}^{Rot}$, causing the resulting $\kappa_{DID}^2$ to be negative. Nevertheless, the values of $\overline{\Delta\alpha_{DID}}$ and $\overline{g_2^{VH}}$ provided for this liquid allows us readily to compute $\overline{R_{VH}}$ at $\lambda_0=633$ nm and $T=293$ K. We find $\overline{R_{VH}}=3.59\times10^{-6}\;\text{cm}^{-1}$, in agreement with the value recently deduced by Pabst \cite{Pabst2022PhD}, viz.  $R_{VH}=3.58\times10^{-6}\;\text{cm}^{-1}$ for this liquid at the same temperature. 

The values of $\overline{R_{VV}}$ we find do not deviate by more than $2\%$ of the experimental values. Since light scattered intensities are measured within $2\%$ relative statistical uncertainty, one would, at first glance, conclude that the theory agrees with experimental data. However, this conclusion is quite rapid : at this stage, one can only conclude that our theoretical findings are consistent with the experimental data that we have used. In order to reach a positive conclusion, one must remember that usually, experimental Rayleigh ratios are determined using the reference sample method \cite{Dezelic1970PureApplChem}. In this method, the Rayleigh ratio for specific incident and scattered wave polarizations (for example $R_{VV}$) must be known with minimal error. Then, using the Hermans-Levinson equation \cite{Hermans1951JOptSoc} which is at the basis of the reference sample method (here applied to $R_{VV}$), viz.
\begin{eqnarray}
    R_{VV}=\left(\frac{\fn}{\fn(\text{ref})}\right)^{2}\frac{I_{VV}}{I_{VV}(\text{ref})}R_{VV}(\text{ref})
    \label{HLEqRVV}
\end{eqnarray}
where $I_{VV}(\text{ref})$, $\fn(\text{ref})$ and $R_{VV}(\text{ref})$ are the measured $VV$ scattered intensity, the refractive index and the $VV$ Rayleigh ratio of the reference, and $I_{VV}$ is the measured scattered intensity of the liquid under consideration for the same wavelength, same temperature and same pressure. Denoting by $I_{VV}^{r}(\text{ref})$ the quantity
\begin{eqnarray}
    I_{VV}^{r}(\text{ref})=\frac{I_{VV}}{I_{VV}(\text{ref})}
    \label{IrVV}
\end{eqnarray}
we can calculate it by changing the reference liquid thanks to the data given in Table \ref{Table4}, and study this ratio as a function of wavelength. The results of these computations are given in Table \ref{Table5}.

\begin{table}[h!]
   \setlength{\tabcolsep}{0.6cm}
   \centering
   \begin{tabular}{ccccccccccccc}
    \hline
    \hline
Liquid & $\lambda_0$ (nm)& $I_{VV}^{r}($C$_6$H$_6$) & $I_{VV}^{r}($C$_7$H$_8$)& $I_{VV}^{r}($CS$_2$) & $I_{VV}^{r}($C$_6$H$_5$NO$_2$)\\
\hline  
CCl$_4$              &   &0.63&0.53&0.2&0.25\\
\\
C$_6$H$_6$           &   &1   &0.85&0.31&0.4\\
\\
C$_7$H$_8$           &546&1.18&1   &0.36&0.5\\
\\
CS$_2$               &   &3.3 &2.79&1   &1.3\\
\\
C$_6$H$_5$NO$_2$     &   &2.53&2.15&0.77&1\\
\hline
CCl$_4$              &   &0.63&0.53&0.2 &0.25\\
\\
C$_6$H$_6$           &   &1   &0.84&0.32&0.4\\
\\
C$_7$H$_8$           &633&1.19&1   &0.38&0.5\\
\\
CS$_2$               &   &3.2 &2.7 & 1  &1.3\\
\\
C$_6$H$_5$NO$_2$     &   &2.4 &2.04&0.77&1\\
\hline
    \end{tabular}
    \caption{$I_{VV}^{r}(\text{ref})$ for two wavelengths at 298 K. The reference liquid is indicated in braces.}
    \label{Table5}
  \end{table} 

Quite remarkably, $I_{VV}^r(\text{ref})$ does depend but little on $\lambda_0$ for a given reference, yet however enough if precision is required if the reference is Benzene or Toluene, particularly for CS$_2$ and C$_6$H$_5$NO$_2$ ; if precision is not fundamental in the sought value of the Rayleigh ratio, then Benzene and Toluene as references are sufficient. Equally remarkably, when the reference is Nitrobenzene,  $I_{VV}^r(\text{ref})$ does not depend on wavelength at all. Computing $I_{VV}^r$(C$_6$H$_6$) for Toluene for $\lambda_0=633$ nm at $T=293$ K with our theoretical values yields $I_{VV}^r$(C$_6$H$_6$)$=1.18$, while Zeissler and Blochowicz experimentally find 1.1  \cite{ZeisslerData2025}. Thus, the theory actually deviates from these experiments by $7\%$ in relative value. 

Since we can compute $I_{VV}^r$(C$_6$H$_6$) for any wavelength at normal pressures in the range $293\leq T\leq300$ K, it is now worthwhile comparing our values with those of Burnham et al.\cite{Burnham75}, which are given for $\lambda_0=514.5$ nm and $T=298$ K. This is shown in Table \ref{Table6}.
\newline
\newline

\begin{table}[h!]
   \setlength{\tabcolsep}{1.35cm}
   \centering
   \begin{tabular}{ccccccccccccc}
    \hline
    \hline
Liquid & $\fn$ &$I_{VV}^{r}($C$_6$H$_6$) & $I_{VV}^{r}($C$_6$H$_6$) \citenum{Burnham75}\\
\hline  
CCl$_4$           &1.462   &0.62&0.64\\
\hline
C$_6$H$_6$        & 1.507  &1   &1\\
\hline
C$_7$H$_8$        & 1.502  &1.18 &1.10\\
\hline
CS$_2$            & 1.638  &3.34 &4.12\\
\hline
C$_6$H$_5$NO$_2$  & 1.562  &2.59 &2.35\\
\hline
    \end{tabular}
    \caption{Our theoretical values of $I_{VV}^{r}($C$_6$H$_6$) vs. those of Burnham et al.\cite{Burnham75} for $\lambda_0=514.5$ nm and $T=298$ K. Values of refractive indices taken from Chang et al.\cite{Chang2024OptExpress}.}
    \label{Table6}
  \end{table}

\noindent
Within the $7\%$ relative deviation alluded to above, those values agree perfectly for CCl$_4$ and C$_7$H$_8$, while our value overestimates that of Burnham et al. by  circa $10\%$ for Nitrobenzene, and underestimates that of Carbon disulfide by circa $21\%$. Here, one may see the effect of  fitting the high-frequency wings to a Lorentzian assimilated to a background, subtracting it from the total spectrum and integrating. For CS$_2$, Battaglia et al. \cite{Battaglia1979MolPhys} found that the part of the signal for CS$_2$ arising from rotational motion represents $58\%$ of the total one, which means that the collision-induced part represents $42\%$ of the total. In our calculations, the DID part for CS$_2$ is represented by $\vert\Psi\vert/(1+\vert\Psi\vert)$ which is $28\%$ for $\lambda_0=546$ nm and circa $25\%$ for $\lambda_0=633$ nm. This is expected for CS$_2$ since its refractive index decreases as $\lambda_0$ increases and the DID effect is controlled by the refractive index as suggested by Eq.\eqref{kappaDID}. Furthermore, before the advent of the laser, the Rayleigh ration of CS$_2$  was deduced for $\lambda_0=436$ nm using this line for a mercury lamp, and $R_u$ was found to lie between $R_u=261\times10^{-6}$cm$^{-1}$ and $R_u=268\times10^{-6}$cm$^{-1}$ at $298$ K \cite{Dezelic1966JCP,Kerker2013Book}. For this wavelength, from Chang et al. \cite{Chang2024OptExpress} we have $\fn=1.6634$ at $295$ K. Using the theoretical value
\begin{eqnarray}
    \overline{R_u}=\overline{R_{VV}}(1+3\overline{r_v})/2
    \label{AvRu}
\end{eqnarray}
together with Eq.\eqref{TheorCompExpRVV} and the values given in Table \ref{Table4} yields $R_u\approx242\times10^{-6}$cm$^{-1}$, which deviates from experiment by $8\%$ \textit{at most}. Since the experimental values of $R_u$ for this liquid at this wavelength were never refined, and that theoretical values were accepted at the time only if they were lower than experimental ones by $10\%$ in relative value at most \cite{Dezelic1970PureApplChem}, our theoretical prediction Eq.\eqref{AvRu} can be accepted as a correct one. Yet, as already mentioned above, this is yet not enough to state that our theoretical findings agree with experimental data in general. A further test of the theory is to check its predictions regarding $(\rho_0\partial_{\rho_0}\fn^2)_T$. As alluded to in the Introduction, Coumou et al. have \textit{experimentally} demonstrated the validity of Eq.\eqref{RisoEinstein} for numerous liquids, by deducing $(\rho_0\partial_{\rho_0}\fn^2)_{T}$ from $\fn$ measurements using an interferometric method and compared Eq.\eqref{RisoEinstein} with Eq. \eqref{Cabannes} for a number of liquids. Dezelic \cite{Dezelic1966JCP} has also provided experimental values of $(\rho_0\partial_{\rho_0}\fn^2)_{T}$. We have ($P$ is pressure in the equation below) \cite{Coumou1964TransFarSoc,Dezelic1970PureApplChem}
\begin{eqnarray}
    \nonumber 
    \left(\rho_0\frac{\partial\fn^2}{\partial\rho_0}\right)_T=\frac{2\fn}{\chi_T}\left(\frac{\partial\fn}{\partial P}\right)_T.
\end{eqnarray}
This shows that obtaining experimental values of $(\rho_0\partial_{\rho_0}\fn^2)_{T}$ requires careful measurements of $\fn$ as a function of pressure $P$ at constant $T$ in order to obtain a reliable experimental value of the derivative in the right hand side of the above equation. The less the liquid is compressible, the more difficult it becomes to precisely obtain this derivative from experimental measurements. Here, we can derive two expressions for $(\rho_0\partial_{\rho_0}\fn^2)_{T}^2$, one for DID-corrected rotation, and one for rotation-corrected DID as follows. We recall that $R_{VV}$, $R_{VH}$ and $R_{iso}$ are related by the model-independent equation \cite{Kielich1965ActaPhysPol,PecoraSteele1965JCP,Dezelic1970PureApplChem}
\begin{eqnarray}
    R_{VV}-\frac{4}{3}R_{VH}=2R_{iso}=\frac{\pi^2}{\rho_0\lambda_0^4}\fS\left(\rho_0\frac{\partial\fn^2}{\partial\rho_0}\right)_T^2,
    \label{RvRhRiso}
\end{eqnarray}
where Eq. \eqref{RisoEinstein} has been used for the second equality. Making use of $R_{VV}$ and $R_{VH}$ derived in Section \ref{5}, $r_v$ given by Eq.\eqref{rvRotZeroDID}, and using also Eq.\eqref{Grot}, we have 
   \begin{eqnarray}
       \left(\rho_0\frac{\partial\fn^2}{\partial\rho_0}\right)_T^2=(\fn^2-1)^2\frac{g_2^{VH}}{G}\left\{1+\frac{4(\fn^2-1)^2(\Delta\alpha_{DID})^2}{9\bar\alpha^2[\fS\left[1+2\kappa(1-g_2^{VH})\right]^2+4\kappa^2g_2^{VH}\left(2-g_2^{VH})\right]}+\cdots\right\}\label{DIDCorrRotEqState}
   \end{eqnarray}
for DID-corrected rotation. If our theory is correct, the above equation must reproduce $(\rho_0\partial_{\rho_0}\fn^2)_{T}$ for C$_6$H$_6$,  C$_7$H$_8$ and CS$_2$ using the values of $\overline{g_2^{VH}}$ and $\overline{\Delta\alpha_{DID}}$ of Table \ref{Table4} in lieu and in place of $g_2^{VH}$ and $\Delta\alpha_{DID}$. Using the same procedure for the rotation-corrected DID version of the theory, we have instead
\begin{eqnarray}
       \left(\rho_0\frac{\partial\fn^2}{\partial\rho_0}\right)_T^2=(\fn^2-1)^2\frac{3-4r_v}{3+16r_v}\left\{1+\frac{4(\fn^2-1)^2(\Delta\alpha_{DID})^2}{9\bar\alpha^2\fS}+\frac{48\kappa^2}{5\fS(3+16r_v)}+\cdots\right\}\label{RotCorrDIDEqState}
   \end{eqnarray}
Eqs.\eqref{DIDCorrRotEqState} and \eqref{RotCorrDIDEqState} can not only be compared with the experimental $(\rho_0\partial_{\rho_0}\fn^2)_{T}$, but also with its expression arising from the Lorenz-Lorentz equation, viz. \cite{Einstein1910AnnPhys}
\begin{eqnarray}
    \left(\rho_0\frac{\partial\fn^2}{\partial\rho_0}\right)_T=\frac{(\fn^2-1)(\fn^2+2)}{3}
    \label{LLindexDerivative}
\end{eqnarray}
which has served many times as an argument for Rayleigh ratio $\fL^4$ scaling. Again, if our theory is correct, then Eq.\eqref{RotCorrDIDEqState} must reproduce $(\rho_0\partial_{\rho_0}\fn^2)_{T}$ for CCl$_4$ and, in principle for  C$_6$H$_5$NO$_2$. Denoting by $A_T$ the theoretical value of $(\rho_0\partial_{\rho_0}\fn^2)_{T}$ predicted by either Eq.\eqref{DIDCorrRotEqState}, either by \eqref{RotCorrDIDEqState}, by $A_E$ its experimental value, and by $A_{LL}$ the Lorenz-Lorentz one given by Eq.\eqref{LLindexDerivative}, the comparison is made in Table \ref{Table7} for $\lambda_0=546$ nm at $T=298$ K and normal pressures.

\begin{table}[h!]
   \setlength{\tabcolsep}{0.6cm}
   \centering
   \begin{tabular}{ccccccccccccc}
    \hline
    \hline
Liquid & $\fn$ & $A_T$ & $A_E$ & $A_{LL}$ & $\delta A_{T-E}$ & $\delta A_{LL-E}$ & $\delta A_{LL-T}$\\
\hline  
CCl$_4$           &1.460   &1.421 & 1.455 & 1.558& $2.3\%$ & $7.1\%$& $9.2\%$\\
\hline
C$_6$H$_6$        & 1.503  & 1.585 &1.655&1.787&$0.9\%$ &$11.7\%$&$12\%$\\
\hline
C$_7$H$_8$        & 1.499  &1.664 &1.6&1.765&$4\%$&$10\%$&$5.9\%$\\
\hline
CS$_2$            & 1.631  & 2.34 &2.37&2.58 & $1.3\%$&$8.8\%$ & $9.7\%$\\
\hline
C$_6$H$_5$NO$_2$  & 1.557  & 2.13 &1.875 &2.10& $13\%$ & $12\%$ & $1.4\%$\\
\hline
    \end{tabular}
    \caption{Theoretical ($A_T$), experimental ($A_E$) and Lorenz-Lorentz ($A_{LL}$) value of $A=(\rho_0\partial_{\rho_0}\fn^2)_{T}$ for $\lambda_{0}=546$ nm and $T$=298 K. $\delta A_{T-E}$, $\delta A_{LL-E}$, $\delta A_{LL-T}$ are respectively relative deviations between the theoretical and experimental value, the Lorenz-Lorentz and the experimental value and the Lorenz-Lorentz and theoretical one . Values of refractive indices taken from Chang et al.\cite{Chang2024OptExpress}.}
    \label{Table7}
  \end{table} 

\end{widetext}
From this last Table, we remark that Eqs. \eqref{DIDCorrRotEqState}, and  \eqref{RotCorrDIDEqState} predict the experimental value of $(\rho_0\partial_{\rho_0}\fn^2)_{T}$ by \textit{at most} $4\%$ save for Nitrobenzene where the deviation is $13\%$. Nevertheless, using the experimental value of $(\rho_0\partial_{\rho_0}\fn^2)_{T}$ quoted in Table \ref{Table7} for C$_6$H$_5$NO$_2$ leads to $R_{iso}=4.01\times10^{-6}$ cm$^{-1}$, which deviates from the value rendered by the Cabannes equation \eqref{Cabannes} by more than $20\%$ (see Table \ref{Table2}). Since there is no obvious reason why Eq.\eqref{RisoEinstein} should fail for this particular liquid, we are lead to the conclusion that for C$_6$H$_5$NO$_2$, the experimental value of $(\rho_0\partial_{\rho_0}\fn^2)_{T}$ is necessarily erroneous, and too small. As already alluded to above, the less compressible the liquid is (and Nitrobenzene has a relatively small $\chi_T$ in regard with other liquids considered here, see Table \ref{Table1}), the more difficult the measurement of $(\rho_0\partial_{\rho_0}\fn^2)_{T}$ becomes. Moreover, quite remarkably, the theoretical value of  $(\rho_0\partial_{\rho_0}\fn^2)_{T}$ and the Lorenz-Lorentz one differ by less than $2\%$, indicating that the Rayleigh ratios for Nitrobenzene \textit{do} scale as $\fL^4$ (in reality, $\fL^4\bar\alpha^4$), and that the main light scattering mechanism at work in this liquid \textit{is} DID. This conclusion agrees with our earlier findings \cite{Dejardin2022PRE2}, where the Kirkwood correlation factor $g_{\rm K}$ was found about unity across a wide temperature range. It follows that light scattering for Nitrobenzene renders information on $\kappa_{DID}>\kappa$, and not $\kappa$, explaining therefore the discrepancy between quantum ab-initio calculations and experimental values of the polarizability anisotropy for this substance \cite{Pabst2022JCP}.  Moreover, whenever rotation dominates DID, our theoretical findings explain why in dilute situations the experimentally determined $\kappa$ does, most of the time, agree with quantum ab-initio calculations.  Finally, let us remark that using Eq.\eqref{OpticalDielectricEqOfState}, we generally have
\begin{eqnarray}
    \left(\rho_0\frac{\partial\fn^2}{\partial\rho_0}\right)_T^2=(\fn^2-1)^2\left[1+\frac{\rho_0}{\fL}\left(\frac{\partial\fL}{\partial\rho_0}\right)_T\right]^2
    \label{GenDer}
\end{eqnarray}
and that Eqs.\eqref{DIDCorrRotEqState}-\eqref{LLindexDerivative} are all of the form of Eq.\eqref{GenDer}. Hence, Eq.\eqref{GenDer} appears as a constraint that $\fL$ is subjected to. This can be perceived as an indication regarding the difficulty in obtaining a general expression for $\fL$ at optical frequencies.

\section{Summary, discussion and conclusion}\label{7}

In this work we have revisited the molecular theory of Rayleigh light scattering by polar fluids in the context of local field theory, confining ourselves to the first Born approximation. This has been accomplished by combining elementary results of radiation theory with a suitable adaptation of electrostatic concepts of local field theory \cite{Bottcher1973Book} to propagation of electromagnetic waves in condensed matter. This has lead us to Eq.\eqref{ElasticRayleighFinal}, which is the starting point of this investigation. We have not shown the derivation steps, and technically, one may suspect from the Keyes-Ladanyi theory that a term coupling the specific part of the local field and rotational terms is missing. Nevertheless, in their simulations of diatomics, Frenkel and McTague \cite{Frenkel1980JCP} have studied this coupling, and concluded that it is negligible. This clearly means that the internal field (i.e., the average local field) is uniform, and as a further result of the first Born approximation (involving only two-body densities in the calculation) this term can be ignored in a classical statistical mechanical theory. Ladanyi has reached the same conclusions two years later \cite{Ladanyi1983JCP} concerning this cross-coupling term. Then, Eq.\eqref{ElasticRayleighFinal} is perfectly justified, differing from Frenkel and McTague's simulation work by a factor $\fL^2$ as far as the rotational term is concerned and Rayleigh scattering involved.  

By restricting ourselves to $90^{\circ}$ light scattering, we have been able to explicit all formal expressions of Section \ref{3}, containing in particular \textit{exact analytical formulas} for the Rayleigh ratios $R_{VV}$ and $R_{VH}$ for gases and liquids made of isotropically polarizable molecules given in Section \ref{3}. The derivation of Rayleigh ratios for pure DID demonstrate that they scale as $\fL^4$ provided $\fS\ll4\kappa_{DID}^2$, and reduce to Einstein's results as DID is ignored (so that $r_v=0$). These formulas were in effect lacking in the theory, leading many authors to conclude that these ratios always scale as $\fL^4$ because of Eq.\eqref{LLindexDerivative}. In effect, CCl$_4$ is the first counterexample to this "rule", since for this liquid $\fS\approx 4\kappa_{DID}^2$, and since $\kappa_{DID}^2$ by its definition \eqref{kappaDID} always scales as $\fL^2$, the scaling of these Rayleigh ratios with the internal field factor $\fL$ is non-trivial, being in fact approximately $\fL^4$ scaling across the liquids we considered for Nitrobenzene only. It was also shown that for Benzene and Toluene, the Rayleigh ratios all scale as $\fL^2$ as the DID contribution is almost negligible in these substances. Those of CS$_2$ are well described by our DID-corrected approximate rotation formulas Eqs.\eqref{RVVDIDCorrRot} and \eqref{RVHDIDCorrRot}, which not only reduce to Einstein's result when $\kappa_{DID}=\kappa=0$, but also to Gans's ones \cite{Gans1921AnnPhys} when DID is completely ignored. We note that the Gans results cannot be obtained from the Rayleigh ratios Eqs.\eqref{RVVDID} and \eqref{RVHHVDID} since $\kappa=0$ there. They cannot be obtained from Eqs.\eqref{RVVRotCorrDID} and \eqref{RVHHVRotCorrDID} either because in the situation described by these equations, DID is the major effect so that $\kappa_{DID}$ \textit{cannot be set equal to zero}. 

We have shown that our theoretical findings can reproduce not only experimental data regarding the Rayleigh ratios of CCl$_4$, C$_6$H$_6$, C$_7$H$_8$, CS$_2$ and C$_6$H$_5$NO$_2$ across a wide wavelength range at room temperature and normal pressures, but is also predictive concerning $\Delta\alpha_{DID}$. In effect, Gabelnick et al.\cite{Gabelnick1968JCP} found a root mean squared DID-induced polarizability anisotropy $\Delta\alpha_{DID}^{RMS}=0.8\;\text{\AA}^3$ for liquid CCl$_4$. Applying the AC peak-to-peak conversion factor $2\sqrt{2}$ to this value yields $\Delta\alpha_{DID}\approx2.13\;\text{\AA}^3$, quite close to ours. Moreover, our feeling that our theoretical findings agree with experimental data is reinforced because they reproduce the experimental values of $(\rho_0\partial_{\rho_0}\fn^2)_T$, hence validating Einstein's result Eq.\eqref{RisoEinstein}, however invalidating the $\fL^4$ scaling rule of Rayleigh ratios. At last, defining a new orientational correlation parameter $g_2^{VH}$ in Fröhlich's manner, we have demonstrated its relation with the older parameter $G$ as determined from the measurement of either $r_v$, either $r_u$. We note in passing that our $G$ value for benzene as rendered by Eq.\eqref{Grot} agrees with Massoulier's ($G\approx 0.56$ while $g_2^{VH}=0.9$) \cite{Massoulier1963JPhys,Massoulier1968CRAS}.     

It is instructive to compute $\kappa_e^2g_2^{BAF}$ and $\kappa^2G$, where $g_2^{BAF}$ is the orientational correlation parameter given by Burnham et al.\cite{Burnham75}, $\kappa_e$ their experimental $\kappa$, and $G$ given by Eq.\eqref{Grot}, yet using the $r_v$ values given by these authors. Without surprise, both values agree fully (i.e. we have $\kappa_e^2g_2^{BAF}=\kappa^2G$). This indicates that the $\kappa_e$ concept is not really necessary. This is so for light scattering experiments as far as for a given molecule, $\kappa$ is well-defined, and molecular rotation dominates the light scattering phenomenon.

Technically, the relation of our theory to the Keyes-Ladanyi one \cite{KeyesLadanyi1984Book} is possible if a) the Kirkwood or Yvon theory of the refractive index \cite{Kirkwood1936JCP,Yvon1937Book} they use is abandoned in favor of Eq.\eqref{OpticalDielectricEqOfState}, b) if the internal field factor $\fL=\fL_L$ in Eq.\eqref{KappaeBL}, if c) $\bar\alpha_e\kappa_e=\bar\alpha\kappa\sqrt{g_2^{VH}}$ and if d) three-body contributions are ignored in the Keyes-Ladanyi theory (since we confine to the first Born approximation, these terms can be ignored). If points a) to d) are fulfilled, the relation between the present theory and that of Keyes and Ladanyi should be $R_{VH}(\text{present work})=\fL_L^{-2}R_{VH}(\text{KL})$, where $R_{VH}(\text{KL})$ is the \textit{total} depolarized Rayleigh ratio of the Keyes-Ladanyi theory \textit{including} the collision-induced term. In fact, the occurrence of $\fL^2$ proportion  between the two theories is not fortuitous, but is a matter of definition. Since in the Keyes-Ladanyi theory the sample occupies the whole space, the \textit{intensity of light scattering} is divided through by the squared Maxwell field amplitude (which then has to be the incident field), rather than that of the vacuum field (which also is the internal field, i.e. the average local field). The proportion between the Maxwell and internal fields is exactly $\fL$ \textit{by definition} \cite{Bottcher1973Book}, independent of the shape of the cavity. It follows that since our theory provides Eq.\eqref{ElasticRayleighFinal}, which does not depend on sample and cavity shapes, that $R_{VH}(\text{KL})$ \textit{does}. Of course, one may argue about the fact that what we just described is a special case, however, this is a clear example, which is enough to demonstrate that $R_{VH}(\text{KL})$ \textit{is not an intensive quantity in general}. Therefore, strictly speaking, $R_{VH}(\text{KL})$ cannot theoretically describe a differential scattering cross-section per unit volume. At last, another feature of our theory is that we provide expressions for $(\rho_0\partial_{\rho_0}\fn^2)_T$, namely, Eqs. \eqref{DIDCorrRotEqState} and \eqref{RotCorrDIDEqState} that \textit{agree} with experimental data, hence \textit{theoretically validating} Eq.\eqref{RisoEinstein}. This cannot be achieved with the present state of development of the Keyes-Ladanyi theory, since nor the Kirkwood, nor the Yvon theory of the refractive index do provide an experimentally valid expression for $(\rho_0\partial_{\rho_0}\fn^2)_T$ \cite{Kerker2013Book,Beysens1977JCP}. Moreover, this means that \textit{it is not because Eq.\eqref{RisoEinstein} holds that Rayleigh ratios should scale as $\fL^4$}. Hence, Eq.\eqref{LLindexDerivative} implies \textit{nothing} regarding the scaling of Rayleigh ratios with $\fL$.

Next, our present work can also be related to Dezelic and Vavra's \cite{DezelicVavra1966CroatChemActa,Dezelic1970PureApplChem} in the following way. Applying Einstein's theory to the off-diagonal elements of the linear dielectric permittivity tensor, it was mentioned in the Introduction that these authors obtained Eq.\eqref{DezelicVavraRan}. Therefore, in the context of Dezelic and Vavra's work, $R_{VH}$ is given by
\begin{eqnarray}
     R_{VH}=\frac{3\pi^2\kappa^2G}{5\rho_0\lambda_0^4}\left(\rho_0\frac{\partial\fn^2}{\partial\rho_0}\right)_T^2\label{DezelicRVH}
\end{eqnarray}
which, because of Eq.\eqref{RisoEinstein} and Eq.\eqref{RvRhRiso}, imply that $R_{VV}$ is given by
\begin{eqnarray}
     R_{VV}&=&\frac{\pi^2}{\rho_0\lambda_0^4}\left[\fS+\frac{4}{5}\kappa^2G\right]\left(\rho_0\frac{\partial\fn^2}{\partial\rho_0}\right)_T^2\label{DezelicRVV}
\end{eqnarray}
and therefore, the depolarization ratio $r_v$ is given by Eq.\eqref{rvG} (hence by Eq.\eqref{rvRotZeroDID} if Eq.\eqref{Grot} is used), leading to the conclusion that $r_v$ does not contain the internal field factor. In contrast, Eqs. \eqref{DezelicRVH} and \eqref{DezelicRVV} do, since $(\rho_0\partial_{\rho_0}\fn^2)_{T}$ contains it due to the optical dielectric equation of state \eqref{OpticalDielectricEqOfState}. Applying now Eq.\eqref{LLindexDerivative}, the Rayleigh ratios Eqs.\eqref{DezelicRVH} and \eqref{DezelicRVV} yields values that are up to $20\%$ \textit{too large} for C$_6$H$_6$, C$_7$H$_8$ and CS$_2$, as already remarked by many authors (including Dezelic and Vavra themselves). For CCl$_4$, they do not work at all since they render $R_{VH}=0$ while for C$_6$H$_5$NO$_2$, they are $8\%$ \textit{too small} (this deviation was accepted as correctness of the theory in these days). Therefore, it follows that if one assumes that Eqs.\eqref{DezelicRVH} and \eqref{DezelicRVV} are correct, the calculation of  $(\rho_0\partial_{\rho_0}\fn^2)_{T}$ from local field theory becomes either a very complicated task, either an ill-posed one for the large majority of liquids, Nitrobenzene being an exception. In effect, as shown here, the precise knowledge of $\fL$ in molecular light scattering theory, and therefore of the precise form of the optical dielectric equation of state, is \textit{unnecessary}. Hence, in reality, the problem of calculating light scattering signals on the basis of obtaining of an explicit form of Eq.\eqref{OpticalDielectricEqOfState} \textit{is an ill-posed one}. This is emphasized by the approximate Eqs.\eqref{DIDCorrRotEqState} and \eqref{RotCorrDIDEqState}, which, if temperature is maintained constant, describe ordinary nonlinear first-order differential equations that, with the help of Eq.\eqref{OpticalDielectricEqOfState}, should allow to approximately determine $\fL$. Here, it fortunately not necessary to accomplish this task. Rather, Eq.\eqref{RvRhRiso} holds since it is a model-independent equation, and if Eq.\eqref{RisoEinstein} is valid, then $(\rho_0\partial_{\rho_0}\fn^2)_T$ in molecular terms can be deduced from this equation if molecular expressions for $R_{VV}$ and $R_{VH}$ are available. Moreover, since $\fL$ and $\bar\alpha$ can formally be eliminated from the Rayleigh ratio scaling, then an equation agreeing with experiment should result for $(\rho_0\partial_{\rho_0}\fn^2)_T$, \textit{without requiring the specification of} $\fL$ in Eq.\eqref{OpticalDielectricEqOfState}. That this is possible is only thanks to the intensive character of the Rayleigh ratios.

Several perspectives can readily be foreseen. First, we may perform studies involving light scattering experiments, simulations and theory in order to be more systematic in analyzing experimental data. This is particularly important concerning, for example, liquid water where $\kappa$ is not so well known. We note that some recent quantum ab-initio studies recently predicted a very small, but non-zero value of the natural polarizability anisotropy of water \cite{Piquemal2007JPhysChem}, while even more recently, the Kerr effect was demonstrated in this liquid \cite{Zalden2018NatureCom} which is a quite unexpected result. In this light, we find it worthwhile investigating again the theory of the electro-optical and optical Kerr effects in dense molecular assemblies in order to establish a relation between the present work and the latter, particularly in the light of including the DID effect in the calculation. This is by no means a trivial problem, particularly when DID is the main mechanism responsible for light scattering. In this situation, only formal expressions are available that are both under-exploited both from the simulation and theoretical points of view.  For example, in liquids such as Nitrobenzene, simulations would be interesting from what precedes because of the results we obtained in Section \ref{6}, despite a strong natural molecular polarizability anisotropy (hence orientational correlations may be relevant only in the electro-optical version where the permanent dipole plays a role \cite{Bottcher1978Book}). Also, molecular liquids such as tri-butyl phosphate do not have a well-defined polarizability anisotropy, which currently forces one to achieve a conformer average in order to extract this quantity from quantum ab-initio calculations \cite{Pabst2022JCP}, making it probably an issue to model Rayleigh ratios in this liquid and similar ones.

Indeed, we cannot close the subject without providing a word regarding dynamic light scattering. Nevertheless, the calculations are fairly more complicated than when DID is not involved. There is no room to discuss the subject in the context of the present work, however, let us mention that due to DID effects, a density-density $G_4-$like function of the kind discussed by Berthier and Biroli \cite{BerthierBiroliRMP2011} is involved in the calculation of the frequency-dependent light scattering signal. From the theoretical point of view, this involves a time-dependent complicated problem that we hope to deal with in a future work.
\newline

\section*{Acknowledgments}
We thank Prof. T. Blochowicz, Dr. F. Pabst and M. R. Zeissler for suggesting the problem, constant support, sharing of some of their light scattering experimental/simulation results and useful and stimulating comments. We are also grateful to Prof. C.Alba-Simionesco, F. Ladieu and Dr. M. Hénot for helpful discussions and useful comments. We furthermore thank Dr. O. Coquand for his help in checking the tedious expressions of the ellipticity factors given in Appendix \ref{AppA}.  At last, we thank Agence Nationale de la Recherche, project reference ANR-24-CE30-0442 (GLASSGO project), for financial support of this work.

\appendix

\section{Explicit expressions for the ellipticity factors $f_{\beps\beps_0}(\kappa)$ for lateral light scattering}\label{AppA}

We provide here expressions for the ellipticity factors arising from DID for lateral light scattering. As far as $g(\br)$ and $W_i$ do not depend on azimuthal angles (see text), these tedious expressions are given by
\begin{widetext}

\begin{eqnarray}
\nonumber
    f_{VV}(\kappa)&=&1+8\kappa\;\indiceGauche{2}{\langle P_{2}(\bu_{1})\rangle}_{0}+16\kappa^{2}\;\indiceGauche{2}{\langle P_{2}(\bu_{1})P_{2}(\bu_{2})\rangle}_{0}\\
    \nonumber
    &+&\frac{(21g_0+3g_2-4g_4)\kappa^{2}}{(7g_0+2g_2+2g_4)}\bigg(1+\indiceGauche{2}{\langle P_{2}(\bu_{1})\rangle}_{0}+\frac{2(7g_0+5g_2+12g_4)}{(21g_0+3g_2-4g_4)}\;\;\indiceGauche{2}{\langle P_{2}^{2}(\bu_{1})\rangle}_{0}\bigg)\\
    \nonumber
    &+&\frac{4(21g_0+3g_2-4g_4)\kappa^{3}}{(7g_0+2g_2+2g_4)}\bigg(\;\indiceGauche{2}{\langle P_{2}(\bu_{1})\rangle}_{0}+\;\indiceGauche{2}{\langle P_{2}(\bu_{1})P_{2}(\bu_2)\rangle}_{0}+\frac{2(7g_0+5g_2+12g_4)}{(21g_0+3g_2-4g_4)}\;\indiceGauche{2}{\langle P_{2}^{2}(\bu_{1})P_{2}(\bu_{2})\rangle}_{0}\bigg)\\
    \nonumber
    &+&\frac{(49g_0-10g_2+4g_4)\kappa^{4}}{2(7g_0+2g_2+2g_4)}\bigg(1+2\;\indiceGauche{2}{\langle P_{2}(\bu_{1})\rangle}_{0}+\;\indiceGauche{2}{\langle P_{2}(\bu_{1})P_{2}(\bu_{2})\rangle}_{0}-\frac{4(7g_0-16g_2+12g_4)}{(49g_0-10g_2+4g_4)}\;\indiceGauche{2}{\langle P_{2}^{2}(\bu_{1})P_{2}(\bu_{2})\rangle}_{0}\\
    &+&\frac{12(7g_0-2g_2+12g_4)}{(49g_0-10g_2+4g_4)}\;\indiceGauche{2}{\langle P_{2}^{2}(\bu_{1})P_{2}^{2}(\bu_{2})\rangle}_{0}-\frac{2(7g_0-32g_2+24g_4)}{(49g_0-10g_2+4g_4)}\;\indiceGauche{2}{\langle P_{2}^{2}(\bu_{1})\rangle}_{0}\bigg)
    \label{f2VV}
\end{eqnarray}

\begin{eqnarray}
\nonumber
    f_{VH}(\kappa)&=&1+2\kappa\;\indiceGauche{2}{\langle P_{2}(\bu_{1})\rangle}_{0}-8\kappa^{2}\;\indiceGauche{2}{\langle P_{2}(\bu_{1})P_{2}(\bu_{2})\rangle}_{0}\\
    \nonumber
    &+&\frac{(98g_0+g_2+8g_4)\kappa^{2}}{(21g_0+3g_2-4g_4)}\bigg(1+\frac{(35g_0-8g_2+20g_4)}{(98g_0+g_2+8g_4)}\indiceGauche{2}{\langle P_{2}(\bu_{1})\rangle}_{0}-\frac{2(14g_0-11g_2+24g_4)}{(98g_0+g_2+8g_4)}\;\;\indiceGauche{2}{\langle P_{2}^{2}(\bu_{1})\rangle}_{0}\bigg)\\
    \nonumber
    &+&\frac{2(49g_0+32g_2+4g_4)\kappa^{3}}{(21g_0+3g_2-4g_4)}\bigg(\;\indiceGauche{2}{\langle P_{2}(\bu_{1})\rangle}_{0}-\frac{7(11g_0-2g_2-4g_4)}{(49g_0+32g_2+4g_4)}\;\indiceGauche{2}{\langle P_{2}(\bu_{1})P_{2}(\bu_2)\rangle}_{0}\\
    \nonumber
    &-&\frac{2(7g_0+26g_2+12g_4)}{(49g_0+32g_2+4g_4)}\;\indiceGauche{2}{\langle P_{2}^{2}(\bu_{1})P_{2}(\bu_{2})\rangle}_{0}\bigg)+\frac{(91g_0-4g_2-4g_4)\kappa^{4}}{(21g_0+3g_2-4g_4)}\bigg(1+\frac{(35g_0+22g_2-20g_4)}{(91g_0-4g_2-4g_4)}\;\indiceGauche{2}{\langle P_{2}(\bu_{1})\rangle}_{0}\\
    \nonumber
    &-&\frac{2(28g_0-13g_2+8g_4)}{(91g_0-4g_2-4g_4)}\;\indiceGauche{2}{\langle P_{2}(\bu_{1})P_{2}(\bu_{2})\rangle}_{0}+\frac{2(35g_0-38g_2+60g_4)}{(91g_0-4g_2-4g_4)}\;\indiceGauche{2}{\langle P_{2}^{2}(\bu_{1})P_{2}(\bu_{2})\rangle}_{0}\\
    &-&\frac{12(7g_0-2g_2+12g_4)}{(91g_0-4g_2-4g_4)}\;\indiceGauche{2}{\langle P_{2}^{2}(\bu_{1})P_{2}^{2}(\bu_{2})\rangle}_{0}+\frac{4(7g_0+5g_2+12g_4)}{(91g_0-4g_2-4g_4)}\;\indiceGauche{2}{\langle P_{2}^{2}(\bu_{1})\rangle}_{0}\bigg)
    \label{f2VH}
\end{eqnarray}
and
 
\begin{eqnarray}
\nonumber
    f_{HH}(\kappa)&=&1-4\kappa\;\indiceGauche{2}{\langle P_{2}(\bu_{1})\rangle}_{0}+4\kappa^{2}\;\indiceGauche{2}{\langle P_{2}(\bu_{1})P_{2}(\bu_{2})\rangle}_{0}\\
    \nonumber
    &+&\frac{(91g_0-4g_2-4g_4)\kappa^{2}}{(21g_0-6g_2+g_4)}\bigg(1-\frac{2(28g_0-13g_2+8g_4)}{(91g_0-4g_2-4g_4)}\indiceGauche{2}{\langle P_{2}(\bu_{1})\rangle}_{0}-\frac{(7g_0-34g_2+22g_4)}{(91g_0-4g_2-4g_4)}\;\;\indiceGauche{2}{\langle P_{2}^{2}(\bu_{1})\rangle}_{0}\bigg)\\
    \nonumber
    &-&\frac{2(91g_0-4g_2-4g_4)\kappa^{3}}{(21g_0-6g_2+g_4)}\bigg(\;\indiceGauche{2}{\langle P_{2}(\bu_{1})\rangle}_{0}+\frac{2(28g_0-13g_2+8g_4)}{(91g_0-4g_2-4g_4)}\;\indiceGauche{2}{\langle P_{2}(\bu_{1})P_{2}(\bu_2)\rangle}_{0}\\
    \nonumber
    &-&\frac{(7g_0-34g_2+22g_4)}{(91g_0-4g_2-4g_4)}\;\indiceGauche{2}{\langle P_{2}^{2}(\bu_{1})P_{2}(\bu_{2})\rangle}_{0}\bigg)+\frac{(189g_0+18g_2+4g_4)\kappa^{4}}{2(21g_0-6g_2+g_4)}\bigg(1-\frac{4(42g_0-15g_2-8g_4)}{(189g_0+18g_2+4g_4)}\;\indiceGauche{2}{\langle P_{2}(\bu_{1})\rangle}_{0}\\
    \nonumber
    &+&\frac{4(21g_0-12g_2+16g_4)}{(189g_0+18g_2+4g_4)}\;\indiceGauche{2}{\langle P_{2}(\bu_{1})P_{2}(\bu_{2})\rangle}_{0}-\frac{8(14g_0-11g_2+24g_4)}{(189g_0+18g_2+4g_4)}\;\indiceGauche{2}{\langle P_{2}^{2}(\bu_{1})P_{2}(\bu_{2})\rangle}_{0}\\
    &+&\frac{(77g_0-26g_2+142g_4)}{(189g_0+18g_2+4g_4)}\;\indiceGauche{2}{\langle P_{2}^{2}(\bu_{1})P_{2}^{2}(\bu_{2})\rangle}_{0}-\frac{4(7g_0+26g_2+12g_4)}{(189g_0+18g_2+4g_4)}\;\indiceGauche{2}{\langle P_{2}^{2}(\bu_{1})\rangle}_{0}\bigg)
    \label{f2HH}
\end{eqnarray}
Further exploitation of these expressions using the rotational mean field approximation may be obtained with the help of
\begin{eqnarray}
    \indiceGauche{2}{\langle P_{2}^{n}(\bu_{1})P_{2}^{m}(\bu_{2})\rangle}_{0}\approx\indiceGauche{1}{\langle P_{2}^{n}(\bu_{1})\rangle}_{0}\indiceGauche{1}{\langle P_{2}^{m}(\bu_{2})\rangle}_{0}
    \label{MeanFieldAv}
\end{eqnarray}
with $m$ and $n$ begin equal to $1$ or $2$ whenever relevant. Because we have $g_{2}^{HH}=g_{2}^{VH}$, in the rotational mean field approximation we have
\begin{eqnarray}
   \indiceGauche{1}{\langle P_{2}(\bu_1)\rangle}_0&=&1-g_{2}^{VH},\label{g2P2}\\
   \indiceGauche{1}{\langle P_{2}^2(\bu_1)\rangle}_0&=&1-\frac{4g_{2}^{VH}}{5},\label{g2P22}
\end{eqnarray}
so that the ellipticity factors can readily be expressed in terms of $g_{2}^{VH}$, $\kappa$, $r_v$ and $\kappa_{DID}$ only.

\end{widetext}


\end{document}